\documentclass[12pt]{iopart}
\usepackage{iopams}
\usepackage{amssymb}
\usepackage{graphicx}
\usepackage{portland}
                       
\def\beq{\begin{equation}}
\def\eeq{\end{equation}}
\def\bea{\begin{eqnarray}}
\def\eea{\end{eqnarray}}
\def\mr{\mathrm}

\def\ug{\,=\,}
\def\lp{\left(}
\def\rp{\right)}
\def\lP{\left[}
\def\rP{\right]}

\def\disp{\displaystyle}

\def\ltord{\hbox{$\;\raise.4ex\hbox{$<$}\kern-.75em\lower.7ex\hbox{$\sim$}
                       \;$}}
\def\gtord{\hbox{$\;\raise.4ex\hbox{$>$}\kern-.75em\lower.7ex\hbox{$\sim$}
                       \;$}}

\begin{document}
\title[Primordial black hole formation in the early universe]
{Primordial black hole formation in the early universe: critical behaviour and 
self-similarity}
\author{Ilia Musco${}^{1}$ and John C. Miller${}^{2,3}$} 

\address{${}^1$Centre of Mathematics for Applications, Department of
Mathematics, University of Oslo, PO Box 1053 Blindern, NO-0316 Oslo, 
Norway\\
${}^2$ Department of Physics (Astrophysics), University of Oxford, Keble
Road, Oxford OX1 3RH, UK\\
${}^3$ SISSA, International School for Advanced Studies, Via Bonomea 265, 
I-34136 Trieste, Italy\\}

\begin{abstract}
 Following on after three previous papers discussing the formation of 
primordial black holes during the radiative era of the early universe, we 
present here a further investigation of the critical nature of the process 
involved, aimed at making contact with some of the basic underlying ideas from 
the literature on critical collapse. We focus on the intermediate state, which 
we have found appearing in cases with perturbations close to the critical 
limit, and examine the connection between this and the similarity solutions 
which play a fundamental role in the standard picture of critical collapse. We 
have derived a set of self-similar equations for the null-slicing form of the 
metric which we are using for our numerical calculations, and have then 
compared the results obtained by integrating these with the ones coming from 
our simulations for collapse of cosmological perturbations within an expanding 
universe. We find that the similarity solution is asymptotically approached in 
a region which grows to cover both the contracting matter and part of the 
semi-void which forms outside it. Our main interest is in the situation 
relevant for primordial black hole formation in the radiative era of the early 
universe, where the relation between the pressure $p$ and the energy density 
$e$ can be reasonably approximated by an expression of the form $p = we$ with 
$w=1/3$. However, we have also looked at other values of $w$, both because 
these have been considered in previous literature and also because they can be 
helpful for giving further insight into situations relevant for primordial 
black hole formation. As in our previous work, we have started our simulations 
with initial supra-horizon scale perturbations of a type which could have come 
from inflation.
 \end{abstract}

\pacs{04.70.-s, 98.80.Cq}

\submitto{\CQG}

\maketitle

\section{Introduction}
 A population of primordial black holes (PBHs) might have been formed in the 
early universe due to gravitational collapse of sufficiently large-amplitude 
cosmological perturbations. Following the initial papers suggesting this 
(Zel'dovich \& Novikov (1969) \cite{Zeldovich}; Hawking (1971) \cite{Hawking}; 
Carr \& Hawking (1974) \cite{Carr1}; and Carr (1975) \cite{Carr2}), various 
authors then investigated the process numerically (Nadezhin, Novikov \& 
Polnarev (1978) \cite{Nadezhin}; Bicknell \& Henriksen (1979) \cite{Bicknell}; 
Novikov \& Polnarev (1980) \cite{Polnarev}; Niemeyer and Jedamzik (1999) 
\cite{Jedamzik1}; Shibata and Sasaki (1999) \cite{Shibata}; Hawke \& Stewart 
(2002) \cite{Hawke}; Musco, Miller \& Rezzolla (2005) \cite{Musco1}). In order 
to form a PBH, the perturbation would need to have an amplitude $\delta$ 
greater than a certain threshold value $\delta_c$ (with $\delta$ often being 
defined as the relative mass excess inside the overdense region, measured at 
the time of horizon crossing). The numerical investigations clarified many 
different aspects of the nature of the process, with particular reference to 
the radiative era of the universe (with the equation of state being taken as 
$p=\frac{1}{3}e$, where $p$ is the pressure and $e$ is the energy density). 
This context represents the most commonly studied scenario for PBH formation.

Niemeyer \& Jedamzik (1999) \cite{Jedamzik1, Jedamzik2} showed that the masses 
of PBHs produced in the radiative era by perturbations of a given profile type, 
follow the typical scaling-law behaviour of critical collapse, first discovered 
for idealized circumstances by Choptuik (1993) \cite{Choptuik}, i.e. the masses 
of the black holes produced follow a power law $M_{BH} \propto (\delta - 
\delta_c)^\gamma$ if $\delta$ is close enough to $\delta_c$. Neilsen \& 
Choptuik (2000) \cite{Choptuik2} later showed that one could get a critical 
collapse, using a succession of ``imploding shells'' of matter as the initial 
conditions, for $p=we$ equations of state with any value of $w$ in the range $0 
- 1$. They found that the value of the critical exponent $\gamma$ was dependent 
only on the value of $w$ and not on the particular form of the perturbation 
profile.

In 2002, Hawke \& Stewart \cite{Hawke} returned to the problem of PBH formation 
in the early universe, with $w = 1/3$, and investigated the nature of the 
collapse going down to smaller values of $(\delta - \delta_c)$ than Niemeyer \& 
Jedamzik had been able to do with their code. For the larger values of $(\delta 
- \delta_c)$, comparable with those of \cite{Jedamzik1}, Hawke \& Stewart again 
found a scaling law with a similar value of $\gamma$, but for smaller $(\delta 
- \delta_c)$ they saw formation of strong shocks and their curve of $\log 
M_{BH}$ against $\log(\delta - \delta_c)$ flattened off at a minimum mass of 
around $10^{-3}$ of their horizon mass. They also found that the value of 
$\delta_c$ depended strongly on the shape of the perturbation, in contrast to 
\cite{Jedamzik1}.
 
In Musco et al (2005) \cite{Musco1}, we considered PBH formation with initial 
conditions given by very small linear perturbations of the energy density and 
velocity fields approximating the growing components of cosmological 
perturbations, and imposed with a length-scale much larger than that of the 
cosmological horizon. The domination of the growing component becomes even 
greater by the time of horizon crossing, as any residual decaying component 
dies away. We again saw a scaling law with similar $\gamma$ for the range of 
values of $(\delta - \delta_c)$ used in \cite{Jedamzik1} but found very 
different values of $\delta_c$ from \cite{Jedamzik1} for similar perturbation 
shapes (we were concentrating on those types of perturbation). This difference 
was attributed to the fact that in \cite{Jedamzik1}, the perturbations were 
made just in the energy density and were imposed directly at the horizon scale, 
so that their value of $\delta_c$ was calculated with inclusion of a 
substantial decaying component which did not then contribute to the black hole 
formation. If one focuses on the effect of perturbations originating from 
inflation, then clearly only growing components will be relevant at much later 
times.

In Polnarev \& Musco (2007) \cite{Musco2}, this kind of cosmologically-relevant 
initial perturbation was imposed in a more precise way, using an asymptotic 
quasi-homogeneous solution \cite{Lifshitz}. Starting from a curvature 
perturbation, which is a time-independent quantity when the perturbation 
length-scale is much larger than the cosmological horizon \cite{Lyth}, 
perturbations in all of the other quantities can then be specified in a 
consistent way, giving a solution with only a growing component. The metric 
perturbation can be large even when the perturbations in energy density and 
velocity are small, and only when there is a large-amplitude metric 
perturbation, corresponding to a non-linear initial perturbation of the 
curvature, can one get $\delta > \delta_c$ at horizon crossing.

In Musco et al (2009) \cite{Musco3}, we returned to the issue raised by Hawke 
\& Stewart \cite{Hawke} concerning whether the scaling law would continue down 
to very small values of $(\delta - \delta_c)$. In order to address this, we 
needed to modify our code with the inclusion of adaptive mesh refinement (AMR) 
so as to be able to handle the extreme conditions which arise near to the 
critical limit. In view of our earlier work, we were particularly wanting to 
investigate the effect of using initial perturbations with just a growing 
component imposed on a scale larger than the horizon, rather than the 
non-linear sub-horizon scale initial perturbations used in \cite{Hawke}. We 
again used the quasi-homogeneous solution to provide our initial conditions. 
Doing this, we did not observe the shock formation seen in \cite{Hawke} (which 
we attributed to the presence of a non-linear decaying component in their 
calculations) but instead found that regular scaling-law behaviour was 
preserved all the way down to the vicinity of the resolution limit of our 
scheme (going beyond the most extreme values shown in \cite{Hawke}). A striking 
feature of our calculations was the appearance of an ``intermediate state'' for 
cases near to the critical limit. For these cases, when the over-dense region 
detaches from the rest of the universe, it reaches a compactness $2M/R \sim 
0.5$ (where $R$ and $M$ are its current radius and mass) at which it then 
remains while proceeding to contract, shedding matter as it shrinks in such a 
way as to maintain $2M/R$ roughly unchanging (with just a small gradual 
decrease). This situation (the ``intermediate state'') persists through many 
e-foldings if $\delta$ is very close to $\delta_c$. Eventually, it reaches a 
mass at which it either enters a final collapse phase leading to black hole 
formation, or disperses into the surrounding medium.

The existence of a similarity solution for cases near to the critical limit, is 
a key feature in the theory of critical collapse \cite{Evans, Gundlach}, and in 
\cite{Musco3} we saw some evidence of self-similarity also in the context of 
PBH formation within an expanding universe, associated with the intermediate 
state. In that paper, however, we were mainly focusing on the issue of 
preservation of the scaling-law. Here, instead, we focus on the issue of 
self-similarity, aiming to clarify the extent to which our intermediate-state 
solution does follow a self-similar behaviour. As a secondary point, we also 
report on calculations which we have made for values of $w$ different from 
$1/3$, using the same approach. Although these other values are of less 
interest from a physical point of view, they have been considered previously in 
the critical-collapse literature, and it can be useful to study them using the 
same approach so as to get further insight into the main case of interest. 
Also, this can be useful for indicating what may happen at epochs of the 
universe at which the equation of state softens due to phase transitions.

For the work of this paper, we have used the same numerical code as in 
\cite{Musco3}, but with some fine tuning of the AMR. Following the present 
Introduction, section 2 reviews our mathematical formulation of the problem and 
presents the self-similar equations as they appear when written in the 
null-slicing foliation used for our numerical calculations. Solutions of the 
self-similar equations are also presented there. Section 3 contains a brief 
summary of the numerical methods used for the simulations. In section 4, we 
make comparison between the simulation results and the similarity solution, and 
section 5 discusses the changes in the values of $\gamma$ and $\delta_c$ which 
come from varying $w$ and the shape of the initial perturbation. Section 6 
contains conclusions. Throughout, we use units for which $c = G = 1$.


\section{Mathematical formulation of the problem}
\label{equations}
\subsection{Cosmic-time slicing}
 For the calculations described here, we have followed the same basic 
methodology as described in our previous papers \cite{Musco1, Musco2, Musco3} 
(which we will refer to as Papers 1, 2 and 3 respectively). We therefore give 
just a brief summary of it here; more details are contained in the previous 
papers.

We use two different formulations of the general relativistic hydrodynamic 
equations: one for setting the initial conditions and the other for studying 
the black hole formation. Throughout, we are assuming spherical symmetry and 
that the medium can be treated as a perfect fluid; we use a Lagrangian 
formulation of the equations with a radial coordinate $r$ which is co-moving 
with the matter.

For setting the initial conditions, it is convenient to use a diagonal form of 
the metric, with the time coordinate $t$ reducing to the standard 
Friedmann-Robertson-Walker (FRW) time in the case of a homogeneous medium with 
no perturbations. (This sort of time coordinate is therefore often referred to 
as ``cosmic time''). We write this metric in the form given by Misner \& Sharp 
\cite{Misner} (but with a change in notation for the metric coefficients):

 \beq 
ds^2=-a^2\,dt^2+b^2\,dr^2+R^2\lp d\theta^2+\sin^2\theta d\varphi^2\rp \, , 
\label{sph_metric}
\eeq
 with $a$, $b$ and $R$ being functions of $r$ and $t$ and with $R$ playing the 
role of an Eulerian radial coordinate. We follow the Misner-Sharp approach for 
writing the GR hydrodynamic equations. Using the notation

\beq
D_t\equiv\frac{1}{a}\lp \frac{\partial}{\partial t}\rp \, ,
\label{D_t}
\eeq
\beq D_r\equiv\frac{1}{b}\lp
\frac{\partial}{\partial r}\rp \, ,
\label{D_r}
\eeq
one defines the quantities

\beq
U \equiv D_t R \, ,
\label{U}
\eeq
and
\beq 
\Gamma \equiv D_r R \, ,
\label{Gamma1}
\eeq
where $U$ is the radial component of four-velocity in the ``Eulerian'' 
frame and $\Gamma$ is a generalized Lorentz factor. The metric coefficient 
$b$ can then be written as

\beq
b \equiv \frac{1}{\Gamma}\frac{\partial R}{\partial r} \, .     
\eeq
 With these specifications, the GR hydrodynamic equations can be written in the 
following form (with the notation that $e$ is the energy density, $p$ is the 
pressure, $\rho$ is the compression factor and $M$ is the mass contained inside 
radius $R$):

 \beq
D_tU=-\lP \frac{\Gamma}{(e+p)}D_rp+\frac{M}{R^2}+4\pi Rp\rP \, ,
\label{Euler1} 
\eeq 
\beq 
D_t\rho=-\frac{\rho}{\Gamma R^2}D_r(R^2U) \, ,
\label{D_trho} 
\eeq 
\beq D_t
e=\frac{e+p}{\rho}D_t\rho \, ,
\label{D_te} 
\eeq 
\beq D_t M=-4\pi R^2 pU \, ,
\label{D_tM} 
\eeq 
\beq D_r a=-\frac{a}{e+p}D_r p \, ,
\label{D_ra} 
\eeq
\beq D_r M=4\pi R^2 \Gamma e \, , 
\label{D_rM}
\eeq 
plus a constraint equation

\beq
\Gamma^2=1+U^2-\frac{2M}{R}\, .
\label{Gamma} 
\eeq
 An equation of state is also needed, and we are here considering ones of the 
form
 
\beq
p = w e \, ,
\label{eq_state}
\eeq
with $w$ being a constant.

\subsection{Initial conditions} 
a perturbation of the otherwise uniform medium representing the cosmological 
background solution, with the length-scale of the perturbation $R_0$ being much 
larger than the cosmological horizon $R_H\equiv H^{-1}$. Under these 
circumstances, the perturbations in $e$ and $U$ can be extremely small while 
still giving a large-amplitude perturbation of the metric (as is necessary if a 
black hole is eventually to be formed) and the above system of equations can 
then be solved analytically to first order in the small parameter $\epsilon 
\equiv (R_H/R_0)^2 \ll 1$. A full discussion of this has been given in Paper 2. 
The result obtained, referred to as the ``quasi-homogeneous solution'', gives 
formulae for the perturbations of all of the metric and hydrodynamical 
quantities in terms only of a curvature perturbation profile $K(r)$, where $r$ 
is the co-moving radial coordinate linked to $R$ by $R = S(t) r$ with $S(t)$ 
being the FRW scale factor; $K(r)$ is conveniently time-independent when 
$\epsilon \ll 1$ \cite{Lyth}.

To characterize the amplitude of the perturbation, we use the integrated 
quantity
 \bea 
\delta(t) = \left( \frac{4}{3}\pi r_0^3 \right)^{-1} \int_0^{r_0} 4\pi r^2 
\left(\frac{e(r,t)-e_b(t)}{e_b(t)}\right) dr \, ,
\label{delta_def}
\eea  
 which measures the relative mass excess within the overdense region, as 
frequently done in the literature; $e_b$ is the background value of the energy 
density and $r_0$ is the co-moving length-scale of the overdense region of the 
perturbation. In Paper 2 (to which we refer for details) it was shown that 

\beq
\frac{e(r,t) - e_b(t)}{e_b(t)}  \equiv \delta e(r,t) = 
\epsilon(t)\frac{3(1+w)}{5+3w}\frac{r_0^2}{3r^2} 
\frac{\partial\lP r^3K(r) \rP}{\partial r} \, ,
\label{de}
\eeq
 and here we consider the initial curvature profile $K(r)$ introduced in Paper 
2\footnote{The equations as presented here include corrections of some 
typographical errors in the equations of Paper 2 which have already been noted 
in Paper 3.}

\beq
K(r) = \lp 1 + \alpha\frac{r^2}{2\Delta^2} \rp \exp \lp -\frac{r^2}{2\Delta^2} 
\rp \, ,
\label{curvature_profile}
\eeq
which implies 

\beq
K^\prime(r) = \frac{r}{\Delta^2} \lP \alpha - \lp1+\alpha\frac{r^2}{2\Delta^2}\rp \rP 
\exp \lp -\frac{r^2}{2\Delta^2} \rp \, .
\eeq
Inserting $K(r)$ and $K^\prime(r)$ into (\ref{de}) one gets

\beq
\hspace{-2.0cm}
\delta e(r,t) = \epsilon(t)\frac{3(1+w)}{5+3w} r_0^2 \lP \lp 1 + \alpha\frac{r^2}{2\Delta^2} \rp
\lp 1 - \frac{r^2}{3\Delta^2} \rp +  \alpha\frac{r^2}{3\Delta^2} \rP \exp \lp -\frac{r^2}{2\Delta^2} \rp .
\label{de2}
\eeq
 The relation between $r_0$ and $\Delta$ for a given $K(r)$, is obtained by 
putting to zero the expression inside the square brackets with $r$ set equal to 
$r_0$. This gives

\beq
r_0^2 = f(\alpha)\Delta^2 \, ,
\label{r_0}
\eeq
where

\bea
f(\alpha) \ug \disp{\left\{
\begin{array}{lc}
\ \,\, 3 \hspace{5.65cm} \mr{if} \quad \alpha = 0 \\ \\
\disp{\frac{(5\alpha-2)+\sqrt{(5\alpha-2)^2+24\alpha}}{2\alpha}} \quad \ \ \mr{if}
\quad \alpha\,\neq 0 \\
\end{array}
\right.} \label{f_alpha}
\eea
 Fixing the parameter $\alpha$ selects a particular perturbation profile, with 
the simplest choice being given by $\alpha=0$, corresponding to a Gaussian 
curvature profile $K(r)$

\beq
K(r) = \exp \lp -\frac{3}{2} \lp \frac{r}{r_0} \rp^{\!2} \rp \, ,
\label{K(r)}
\eeq
 which gives a Mexican hat profile for the energy density perturbation $\delta 
e(r,t)$, as often used in in the literature \cite{Jedamzik1,Musco1,Musco3}:

\beq
\delta e(r,t)  \ug  \epsilon(t)\frac{9(1+w)}{5+3w} \Delta^2 \lp 1 \,-\,\frac{r^2}{3\Delta^2} \rp
\exp{\lp -\frac{r^2}{2\Delta^2} \rp} \, .  
\label{mex_hat} 
\eeq
 The perturbation shape is fixed by the choice of $\alpha$, while the 
perturbation amplitude is determined by the value chosen for $\Delta$. 
Inserting (\ref{de2}) into (\ref{delta_def}) one gets

\beq
\delta(t) =  \epsilon(t) \frac{3(1+w)}{5+3w}f(\alpha)\Delta^2 \lp 1+\alpha\frac{f(\alpha)}{2} \rp 
\exp{\lp -\frac{f(\alpha)}{2}\rp} \, ,
\label{delta}
\eeq
 within the linear regime, where $\epsilon \ll 1$, with the time evolution of 
the perturbation being given by $\epsilon(t)$, which measures how the 
perturbation length-scale $R_0$ changes with respect to the cosmological 
horizon scale $R_H$. The expression for $\epsilon(t)=(R_H/R_0)^2$ involves 
only background quantities for the unperturbed universe and inserting this 
gives
 \beq
\delta(t) = \delta(t_i) \lp \frac{t}{t_i} \rp^\frac{2(1+3w)}{3(1+w)}\, ,
\eeq
 which leads to the familiar relationships $\delta(t) \propto t$ when $w=1/3$ 
and $\delta(t) \propto t^{2/3}$ when $w = 0$. 

For our discussion here, we need a value for $\delta$ measured in a standard 
way, and it is convenient to use for this the value given by taking 
$\epsilon(t) = 1$ in (\ref{delta}). For small perturbations, this value for 
$\delta$ coincides with the one at the horizon-crossing time, and it does not 
differ by very much from that even for the larger perturbations of interest for 
PBH formation, at least for the types of perturbation profile being used here. 
Since evaluating (\ref{delta}) analytically in this way is much more precise 
than making a numerical integral on the grid at horizon-crossing time, we use 
this value here, as we did in Papers 2 and 3.

\subsection{Null slicing}
\label{hernandez}
 In our work, the Misner-Sharp approach, with cosmic time slicing, has been 
used for setting the initial conditions and then for evolving them to produce 
corresponding initial conditions on an outgoing null slice. These are then 
passed to the main code which uses the ``observer time'' null-slicing 
formulation of Hernandez \& Misner \cite{Hernandez} for following the further 
evolution leading up to black hole formation. In this formulation, each 
outgoing null slice is labelled with a time coordinate $u$, which takes a 
constant value everywhere on the slice, and the metric (\ref{sph_metric}) is 
re-written as 
 \beq 
ds^2=-f^2\,du^2-2fb\,dr\,du+R^2\left(d\theta^2+\sin^2\theta 
d\varphi^2\right) \, , 
\label{nullmetric} 
 \eeq
where $f$ is the new lapse function which determines how the value of $u$ 
changes from one slice to another. The lapse $f$ needs to be normalised by 
setting $u$ equal to the proper time of some suitable observer, and the 
standard way of doing this is to equate it to the proper time of a distant 
observer, as we have done in our previous papers (setting $f = 1$ at the 
location of that observer). However, this is not suitable for our present 
discussion of the connection with similarity solutions, because distant 
observers are not within the region covered by the similarity solution. Because 
of this, we instead synchronise here with the proper time of a central 
observer, setting $f = 1$ at $r = 0$.

Within this formulation, the operators equivalent to (\ref{D_t}) and 
(\ref{D_r}) are
 \beq
D_t \equiv \frac{1}{f}\left(\frac{\partial}{\partial u}\right) \, ,
\label{D_u}
\eeq
\beq
D_k \equiv \frac{1}{b}\left(\frac{\partial}{\partial r}\right) \, ,
\label{D_k}
\eeq
where $D_k$ is the radial derivative in the null slice and the 
corresponding derivative in the Misner-Sharp space-like slice is given by
 \beq
D_r=D_k-D_t \, .
\label{D_r_2}
\eeq
The hydrodynamic equations can then be formulated in a way analogous to what 
was done in cosmic time. The system of equations used in the main code, 
replacing the cosmic-time ones (\ref{Euler1})  -- (\ref{D_rM}), are:
 \beq
\hspace{-1.5cm}
D_tU=-\frac{1} {1-c_s^2}\left[\frac{\Gamma}{(e+p)}D_kp + \frac{M}{R^2} 
+ 4\pi Rp +  c_s^2\left(D_kU+\frac{2U\Gamma}{R}\right)\right], 
\label{Euler2}
\eeq
\beq
D_t\rho=\frac{\rho}{\Gamma}\left[D_tU-D_kU-\frac{2U\Gamma}{R}\right] \, ,
\label{D_trho2}
\eeq
\beq
D_te=\left(\frac{e+p}{\rho}\right)\,D_t\rho \, ,
\label{D_te2}
\eeq
\beq
D_k\left[\frac{(\Gamma + U)}{f}\right] = -4\pi R \lp \frac{e+p}{f} \rp \, ,
\label{fnew}
\eeq
\beq
D_t M=-4\pi R^2 pU \, ,
\label{D_tM2}
 \eeq
\beq
D_kM=4\pi R^2[e\Gamma-pU],
\label{D_kM}
\eeq
 where $c_s = \sqrt{(\partial p/\partial e)}$ is the sound speed.\footnote{Note 
that there was a typographical error in equation (27) of Paper 1; equation 
(\ref{fnew}) here is the correct form which has been used throughout for the 
calculations in all of our papers.} The quantity $\Gamma$ is given by 
(\ref{Gamma}), as before, and we also have
 \beq   
\Gamma=D_kR - U \, ,  
\eeq
which replaces equation (\ref{Gamma1}) in this slicing.

\subsection{Self-similarity: equations and solution}
\label{similarity}
 In this paper, we are wanting to investigate the connection between our 
numerical calculations for gravitational collapse of cosmological perturbations 
within the expanding medium of the early universe, and studies by previous 
authors looking at the problem of critical collapse under simpler circumstances 
(in particular, with asymptotic flatness) where the presence of similarity 
solutions plays a key role. There, as $(\delta - \delta_c) \to 0$, one 
approaches a critical solution where all of the matter in the original 
contracting region is shed during the contraction which ends, with zero matter, 
at a time referred to as the critical time $t_c$. The later stages of this 
follow a similarity solution. For small positive values of $(\delta - 
\delta_c)$, the similarity solution is closely approached but eventually there 
is a divergence away from it, with the remaining material collapsing to form a 
black hole. In our previous calculations, we have noted the appearance of an 
``intermediate state'' during which the behaviour gave some indication of being 
roughly self-similar. Here, we are wanting to investigate this in more detail 
in order to understand how closely this reproduces what was seen in the 
previous critical collapse work. For doing this, we first needed to derive the 
equations governing the similarity solutions in the Hernandez-Misner foliation, 
as used for our numerical calculations.

Our self-similar coordinate $\xi$ is the ratio between the distance away from 
the centre of symmetry and the time away from the critical time, both of which 
are given in terms of invariantly-defined quantities. For the distance, we use 
the circumferential radius $R$ and for the time we use the outgoing null time 
$u$ with the lapse $f$ set equal to $1$ at the centre, so that $u$ is 
synchronized with the proper time of a central observer. (Note that this 
approach is importantly different from that of Evans \& Coleman \cite{Evans} 
who normalised the time differently and obtained a more complicated form for 
their similarity solution.) For deriving the hydrodynamical equations in 
self-similar form, it is convenient to set the zero point of the time scale to 
be at the critical time. The times being considered are ones before the 
critical time and hence take negative values; we then define our self-similar 
coordinate, which needs to be positive, as
 \beq
\xi \equiv - \frac{R}{u} \, ,
\label{xi_ot}
\eeq
 and look for a solution of the system of partial differential equations of the 
previous section in terms of quantities depending only on $\xi$. For that set 
of quantities we have $U$, $\Gamma$ and $f$, as already defined, and two new 
ones: $\Omega \equiv 4\pi R^2e$ and $\Phi \equiv M/R$, while we eliminate 
$\rho$ by combining equations (\ref{D_trho2}) and (\ref{D_te2}), and write the 
pressure $p$ in terms of the energy density $e$ using the equation of state $p 
= we$ (which gives the sound-speed $c_s = \sqrt{w}$). To derive the 
self-similar form of the equations, we then proceed as follows. First, we write 
the Hernandez-Misner equations in an ``Eulerian'' form, using coordinates $R$ 
and $u$. For a general quantity $Y$, we have
 \beq
dY = \left(\frac{\partial Y}{\partial u}\right)_{\!R}\,du + 
\left(\frac{\partial Y}{\partial R}\right)_{\!u}\,dR \, ,
\eeq
 which then gives
 \bea
\left(\frac{\partial Y}{\partial u}\right)_{\!r} & = & 
\left(\frac{\partial Y}{\partial u}\right)_{\!R}  
+ \left(\frac{\partial Y}{\partial R}\right)_{\!u} 
\left(\frac{\partial R}{\partial u}\right)_{\!r} \, , \nonumber \\
& \to & \left(\frac{\partial Y}{\partial u}\right)_{\!R}      
+ fU\left(\frac{\partial Y}{\partial R}\right)_{\!u} \, ,
\eea
 inserting the expression for $U \equiv D_t\,R$. Similarly
 \bea
\left(\frac{\partial Y}{\partial r}\right)_{\!u} & = &  
\left(\frac{\partial Y}{\partial R}\right)_{\!u} 
\left(\frac{\partial R}{\partial r}\right)_{\!u} \, , \nonumber \\
& \to & b\,(\Gamma + U)\left(\frac{\partial Y}{\partial R}\right)_{\!u}   
\eea                                        
 inserting the expression for $\Gamma \equiv D_r\, R$. The operators $D_t$ and 
$D_k$ can therefore be written in the Eulerian form
\bea
D_t \equiv \frac{1}{f}\left(\frac{\partial}{\partial u}\right)_{\!r}
= \frac{1}{f}\left(\frac{\partial}{\partial u}\right)_{\!R}
+ U\left(\frac{\partial}{\partial R}\right)_{\!u} \, , \\
D_k \equiv \frac{1}{b}\left(\frac{\partial}{\partial r}\right)_{\!u}
= (\Gamma + U)\left(\frac{\partial}{\partial R}\right)_{\!u} ,
\eea
which, combined with the definition of the self-similar coordinate $\xi$, gives
\bea
D_t = \frac{1}{R}\left( U + \frac{\xi}{f} \right) \frac{d}{d\ln\xi} \, , \\
D_k = \frac{1}{R}(\Gamma + U)\frac{d}{d\ln\xi} ,
\eea
 One then finds that equations (\ref{Gamma}), (\ref{Euler2}) - (\ref{D_kM}) can 
indeed be recast as a system for $U$, $\Omega$, $\Phi$, $f$ and $\Gamma$, 
depending only on $\xi$. After a considerable amount of algebra, we arrive at 
a set of three ordinary differential equations
 \bea
\frac{d\ln U}{d\ln\xi} = \lP \frac{(\Phi+w\Omega)^2 - 2w\Gamma^2\Phi}{U^2(\Phi+w\Omega)^2 - 
w\Gamma^2(\Omega-\Phi)^2} \rP \lP (\Omega-\Phi) - \frac{(1+w)\Omega U}{(\Gamma+U)} \rP 
\label{self_U_ot} \\
& \nonumber \\
\frac{d\ln \Omega}{d\ln\xi} =  \frac{(1+w)(\Omega-\Phi)}{(\Phi+w\Omega)}\frac{d\ln U}{d\ln\xi} + 
\frac{2w}{(\Phi+w\Omega)}  \lP (\Omega-\Phi) - \frac{(1+w)\Omega U}{(\Gamma+U)} \rP 
\label{self_Omega_ot}\\
& \nonumber \\
\frac{d\ln \Phi}{d\ln\xi} = \frac{1}{\Phi} \lP (\Omega-\Phi) - \frac{(1+w)\Omega U}{(\Gamma+U)} \rP 
\label{self_Phi_ot} \
\eea
together with two algebraic equations 
\bea
\Gamma = 1 + U^2 -2\Phi \,, \\
f = - \frac{\xi}{(1+w)\Omega U} \lP (\Omega-\Phi) - 
\frac{U}{\Gamma}(\Phi+w\Omega) \rP \, .
\eea 
 Note that the three ODEs do not depend explicitly on the lapse $f$. However, 
$U$, $\Omega$ and $\Phi$ all require a boundary condition. Clearly they all go 
to zero at $R=0$, and hence at $\xi=0$, but since the equations involve 
derivatives with respect to $\ln\xi$, we need to series-expand away from 
$\xi=0$ in order to start our integration. To get the expression for $U$, it is 
necessary to specify the central value of $f$ which, as already mentioned, we 
are setting equal to $1$. For small $\xi$, we then find
 \bea
\disp{\left\{
\begin{array}{ll}
& U(\xi) = - \disp{\frac{2}{3(1+w)}} \xi \\
& \\
& \Phi(\xi) = k \xi^2 \\
& \\
& \Omega(\xi) = 3 k \xi^2 
\end{array}
\label{bc}
\right.}
\eea
 to lowest order in $\xi$, with the constant $k$ being determined by requiring 
regularity of the solution for $U$ at the critical point, where the numerator 
and denominator of the first term on the right hand side of equation 
(\ref{self_U_ot}) both go to zero.

\begin{figure}[t!]
\centering
\includegraphics[width=7cm]{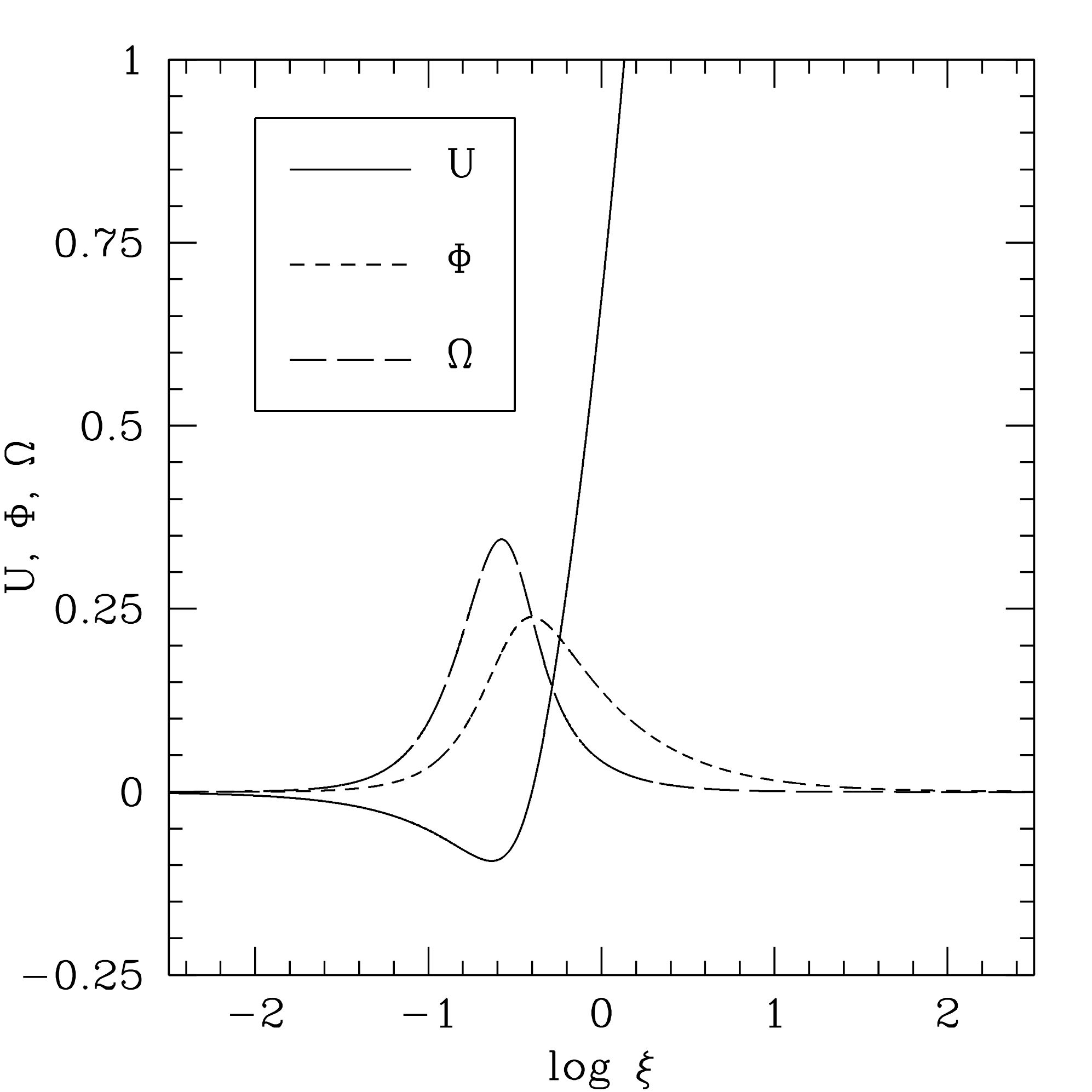}
 \caption{\label{fig.1} \small The self-similar solutions in null time for 
$U$, $\Phi$ and $\Omega$ plotted against $\log \xi$ for $w = 1/3$.}
 \end{figure}

The integration of equations (\ref{self_U_ot}) - (\ref{self_Phi_ot}) was 
carried out using a fourth-order Runge-Kutta scheme, with a shooting method 
being used to determine the value of $k$. Since $U$ can be negative, the 
derivative $d\ln U/d\ln\xi$ needs to be re-written as $(1/U)\,dU/d\ln\xi$ when 
performing the integration. Results are shown in figure \ref{fig.1} where $U$, 
$\Omega$ and $\Phi$ are plotted as functions of $\log \xi$ for $w = 1/3$.



\section{The method used for the numerical simulations} 
\label{method} 
 The present calculations for PBH formation have been made with the same code 
as used in Paper 3. Since this has been fully described previously, we will 
just give a brief outline of it here. It is an explicit Lagrangian 
hydrodynamics code based on that of Miller \& Motta (1989) \cite{Miller1} but 
with the grid organized in a way similar to that of Miller \& Rezzolla (1995) 
\cite{Miller2} which was designed for calculations in an expanding cosmological 
background. The code has a long history and has been carefully tested in its 
various forms. The basic grid uses logarithmic spacing in a mass-type comoving 
coordinate, allowing it to reach out to very large radii while giving finer 
resolution at small radii. Our initial data is derived from the 
quasi-homogeneous solution and is specified on a space-like slice (at constant 
cosmic time) with $\epsilon = 10^{-2}$, giving $R_0 = 10\,R_H$. The outer edge 
of the grid has here been placed at $1000\,R_H$ for convenience in making some 
of the plots; putting it at $90\,R_H$, as we did previously, was already 
sufficient for ensuring that there is no causal contact between it and the 
perturbed region during the time of the calculations. The initial data is then 
evolved using the Misner-Sharp equations (\ref{Euler1}-\ref{Gamma}), so as to 
generate a second set of initial data on a null slice and the null-slice 
initial data is then evolved using the Hernandez-Misner equations (see 
\cite{Musco1}).

For the calculations presented in Paper 3, we introduced an adaptive mesh 
refinement scheme (AMR), on top of the existing logarithmic grid, giving us 
sufficient resolution so as to be able to follow black hole formation down to 
extremely small values of $(\delta - \delta_c)$. Having the AMR is particularly 
important for allowing us to follow the deep voids which form outside the 
central contracting region in cases very close to the critical limit. The same 
scheme has been used with just minor modifications for the calculations 
presented here. Our aim in writing the AMR was to avoid the use of artificial 
viscosity, but it has now emerged that some residual artificial viscosity was 
still present and is necessary for correct functioning of the code in its 
present form. The presence of this is not thought to affect the results 
presented in any important way, however. We have successfully used the scheme 
with more than thirty levels of refinement and all relevant features of the 
solutions have been fully resolved.



\section{Comparison between the simulation results and the similarity 
solution}
 In this section, we present results from our investigation of the extent to 
which the ``intermediate state'' seen previously in our numerical simulations 
corresponds to the similarity solution discussed in section \ref{similarity}. 
First, we recall the background to this coming from our earlier work.

At the beginning of our calculations, we have a growing perturbation, with a 
length-scale larger than the cosmological horizon, consisting of a slight 
over-density which is expanding along with the rest of the universe but a 
little more slowly. After it enters within the cosmological horizon, it starts 
to contract and then collapse if its amplitude $\delta$ is large enough. As 
described earlier, if $\delta$ is greater than the critical value $\delta_c$, 
it goes on to produce a black hole, while for smaller values it eventually 
disperses into the background medium. Initially, the over-dense region has 
decreasing compactness $2M/R$ but, if $\delta$ is only slightly greater than 
$\delta_c$, it then reaches a value of $2M/R$ at which it remains with very 
little change as it proceeds to contract, shedding matter as it shrinks in such 
a way as to keep $2M/R$ roughly constant (but with just a small continuing 
gradual decrease). For a radiation fluid with $w = 1/3$, this ``intermediate 
state'' has $2M/R \sim 0.5$, quite far from the condition for a black hole. A 
large pressure gradient develops at the edge of the contracting matter and it 
is this which drives the wind, opening up a deep void between the contracting 
matter and the rest of the universe. Eventually, there is a deviation away from 
the intermediate state and $2M/R$ increases towards $1$. This process was 
investigated in detail in Paper 3 for a radiation fluid with $w = 1/3$ and some 
evidence was seen of self-similar behaviour during the period of the 
intermediate state. We investigate this further here. Throughout this section, 
we consider only $w = 1/3$ and, since we are dealing with an astrophysical 
application, we use $t$ to denote time even when we are dealing with the null 
time previously denoted by $u$.

\begin{figure}[t!]
\centering
\includegraphics[width=7cm]{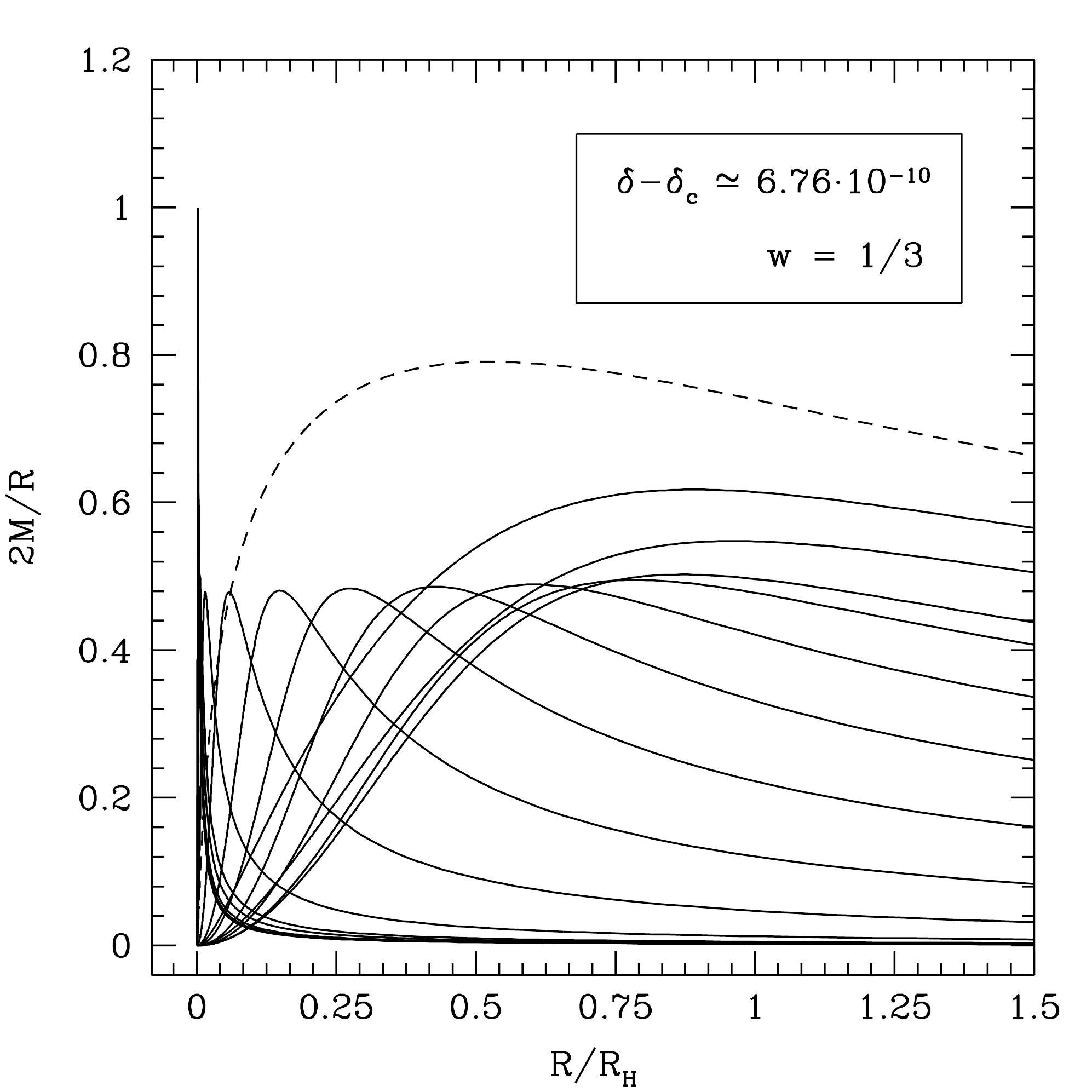}
\includegraphics[width=7cm]{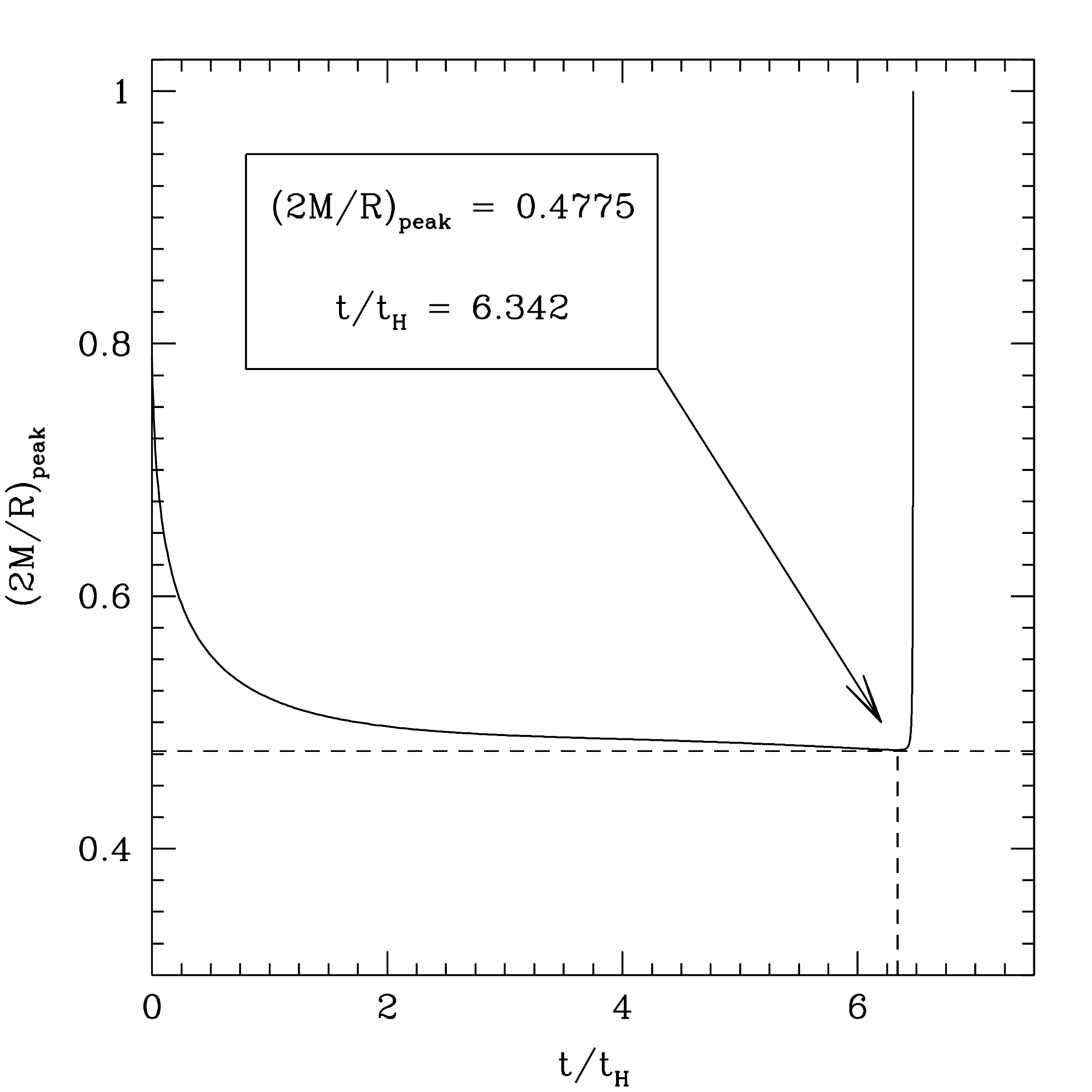}
 \caption{\label{fig.2} \small In the left-hand frame, we show the behaviour of 
$2M/R$ for a nearly critical case plotted against $R/R_H$ at different time 
levels, where $R_H$ is the cosmological horizon scale at the moment of horizon 
crossing. The dashed curve indicates the initial conditions used by the 
null-time code. The right-hand frame shows the time evolution of the peak of 
$2M/R$ during the intermediate state with $t/t_H$ being the null time 
(normalised in the same way as for the similarity solution) measured in units 
of the horizon crossing time $t_H$ ($=R_H/2$). The horizontal dashed line 
indicates the value of $(2M/R)_{\rm peak}$ coming from the corresponding 
similarity solution, while the vertical dashed line indicates when the 
intermediate state ends for this case, very close to the critical time. In the 
collapse following this, $(2M/R)_{\rm peak}$ increases rapidly towards $1$ with 
the formation of the black hole.}
 \end{figure}

For making our discussion, we focus on a particular case near to the critical 
threshold, starting from a standard Mexican hat perturbation in the energy 
density (i.e. $\alpha = 0$) with $(\delta - \delta_c) = 6.76 \times 10^{-10}$. 
This gives rise to a black hole with mass $2.14 \times 10^{-3}\,M_H$, where 
$M_H$ is the cosmological horizon mass at the horizon-crossing time. The 
left-hand frame of figure \ref{fig.2} shows the behaviour of $2M/R$ plotted 
against $R/R_H$ at different time levels for this case ($R_H$ being the 
cosmological horizon scale at the moment of horizon crossing). The dashed curve 
shows the initial conditions used by the null-time code. The subsequent 
evolution proceeds in the direction of decreasing $(2M/R)_{\rm peak}$ until the 
intermediate state is reached, and then the peak moves progressively inwards. 
When $\delta$ is extremely close to the critical limit, as it is here, the 
intermediate state with almost constant $(2M/R)_{\rm peak}$, persists through 
many e-foldings and the eventual turn-off away from it, with $(2M/R)_{\rm 
peak}$ rising towards the black-hole value of $1$, happens very abruptly.

\begin{figure}[t!]
\centering
\includegraphics[width=7cm]{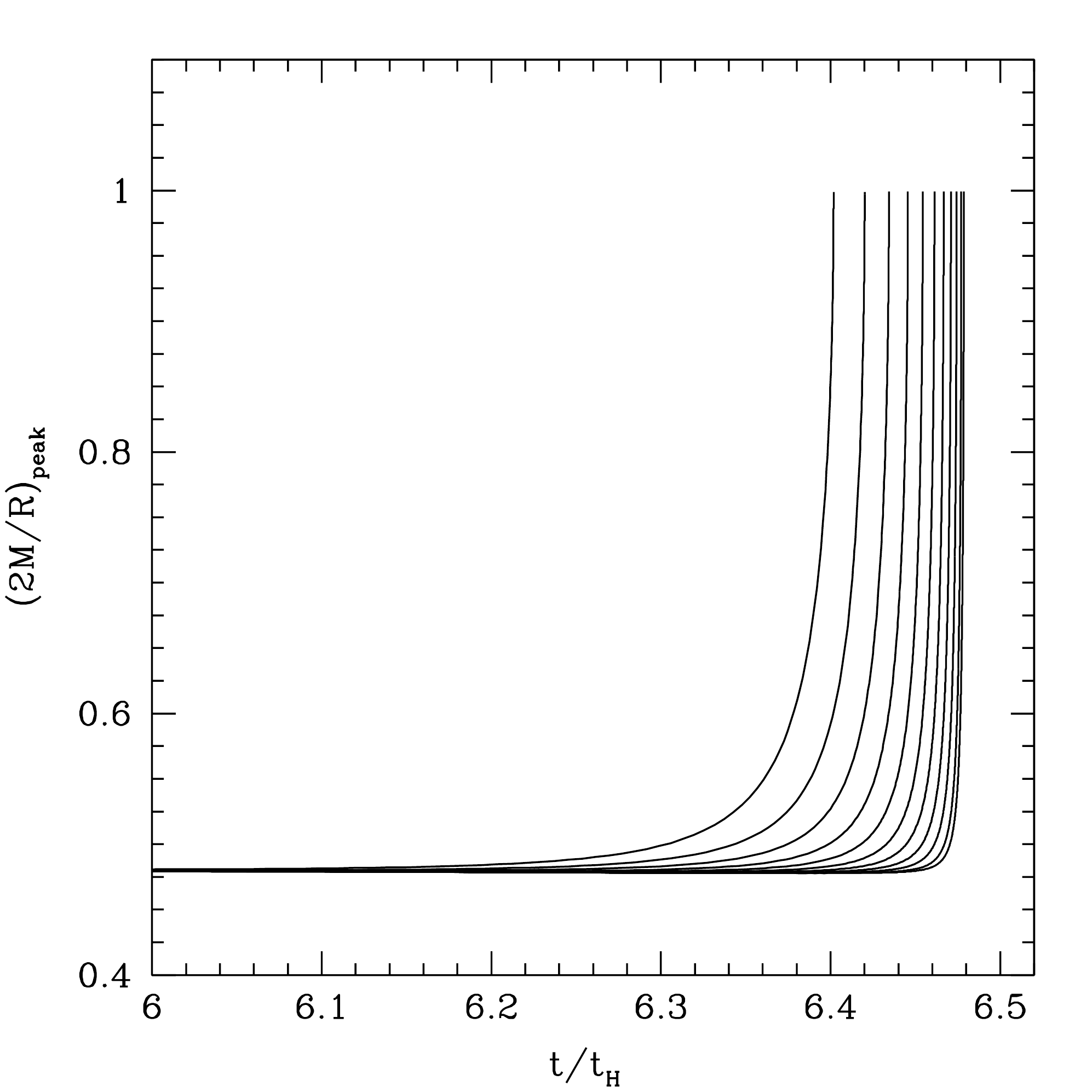}
\includegraphics[width=7cm]{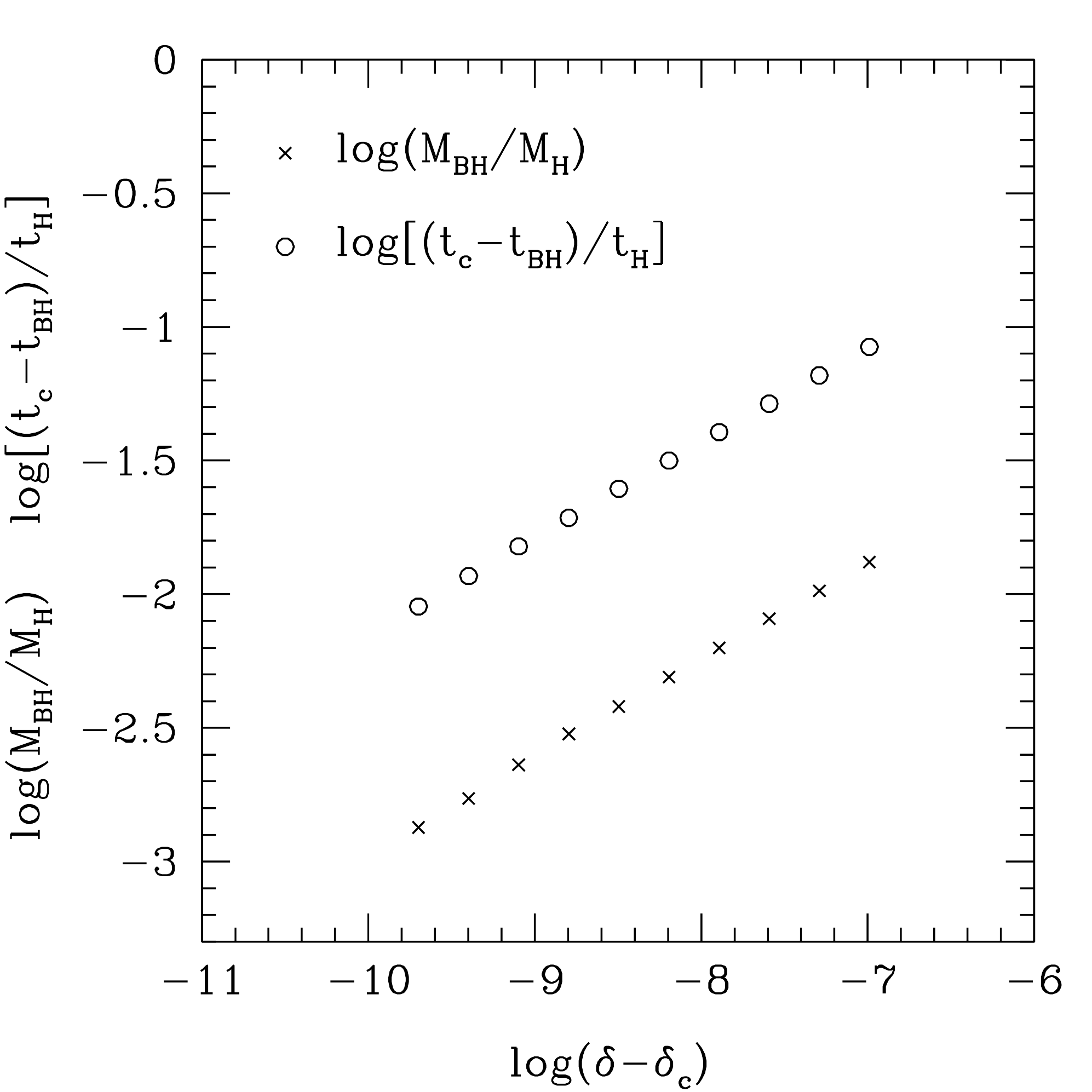}
 \caption{\label{fig.3} \small In the left-hand frame, $(2M/R)_{\rm peak}$ is 
plotted as a function of time for a succession of values of $(\delta-\delta_c)$ 
equally spaced in the log, showing the convergence of the black-hole formation 
time $t_{BH}$ as $(\delta-\delta_c)$ is reduced. The right-hand frame shows 
the scaling-law behaviour of $M_{BH}/M_H$ and $(t_c - t_{BH})/t_H$ as a 
function of $(\delta-\delta_c)$.}
 \end{figure}

Another view of this behaviour is shown in the right-hand frame of figure 
\ref{fig.2}, where the time evolution of $(2M/R)_{\rm peak}$ is plotted, from 
the same run as shown in the left-hand frame, again starting the plot from the 
time of the initial data for the null-time code. The time coordinate used is 
normalised as described above. The horizontal dashed line shows the value of 
$(2M/R)_{\rm peak}$ coming from the similarity solution (corresponding to the 
peak of $\Phi$ in figure \ref{fig.1} -- note that $(2M/R) = 2\Phi$). In the 
simulation, the peak starts from a fairly high value after the perturbation has 
re-entered the cosmological horizon, and then decreases while the overdense 
region continues expanding along with the rest of the universe. Eventually the 
expansion of the overdense region reverses, and $(2M/R)_{\rm peak}$ settles 
into the roughly-constant value of the intermediate state, approaching the 
similarity value more and more closely as time goes on. It is almost touching 
the similarity line at the moment when the intermediate state ends and the 
rapid collapse towards black-hole formation begins. The behaviour is roughly 
similar for the other values of $w$ which we have tested (in connection with 
the discussion in the next section). The intermediate state lasts for longer 
with smaller values of $w$ and the similarity value of $(2M/R)_{\rm peak}$ 
changes with $w$ in such a way as to remain roughly equal to $\delta_c$.

In figure \ref{fig.3}, the left-hand frame is a zoom of a series of curves 
similar to that in the right-hand frame of figure \ref{fig.2}, but drawn for a 
succession of values of $(\delta - \delta_c)$, equally-spaced in the log, 
decreasing from left to right. It can be seen that, as $(\delta - \delta_c)$ 
decreases, the rising part of the curve becomes progressively more abrupt and 
the value of $t$ when $(2M/R)_{\rm peak} \to 1$ (which we will call $t_{BH}$) 
is tending towards a limiting value $t_c$ as $\delta \to \delta_c$. This is the 
critical time needed for comparing with the similarity solutions, with $\xi$ 
for the simulation results set equal to $R/(t_c - t)$. As can be seen from the 
right-hand frame, $(t_c - t_{BH})$ again follows a scaling law in $(\delta - 
\delta_c)$ which has the same value of the exponent as for the black hole 
masses (also plotted here).

If the similarity solution were being followed exactly, the central energy 
density would scale with time as $(t_c - t)^{-2}$ (this follows from the 
behaviour of $\Omega$ given in equation (\ref{bc})). An alternative way of 
obtaining a value for $t_c$ is therefore to fit this scaling to the simulation 
results. This is less precise than the limiting procedure discussed above 
(because the similarity solution is only being followed approximately) but the 
value obtained for $t_c$ in this way agrees with the previous one to around one 
part in $10^4$ if the time-variation at the closest approach to the similarity 
solution is used.

\begin{figure}[t!]
\centering
\includegraphics[width=7cm]{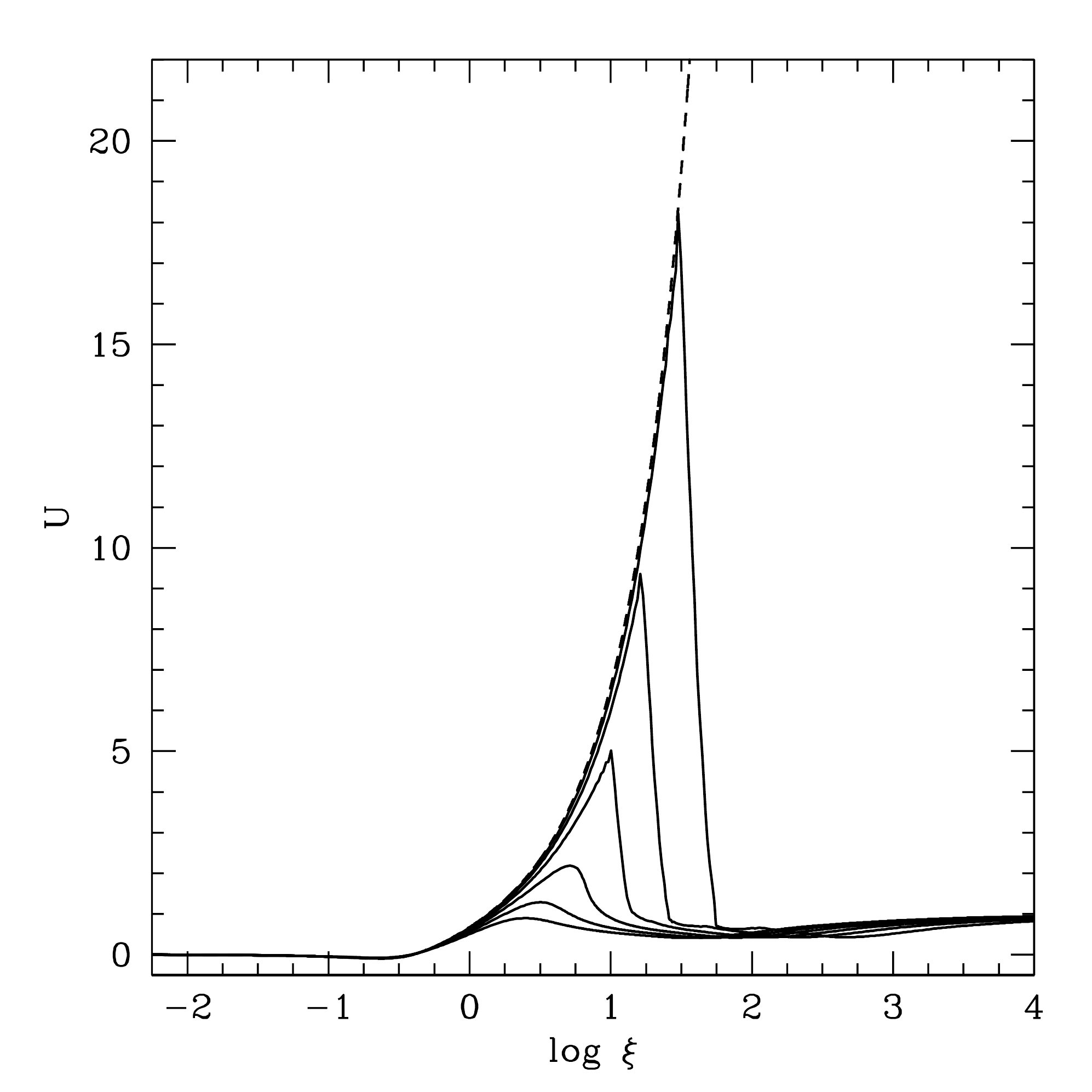}
\includegraphics[width=7cm]{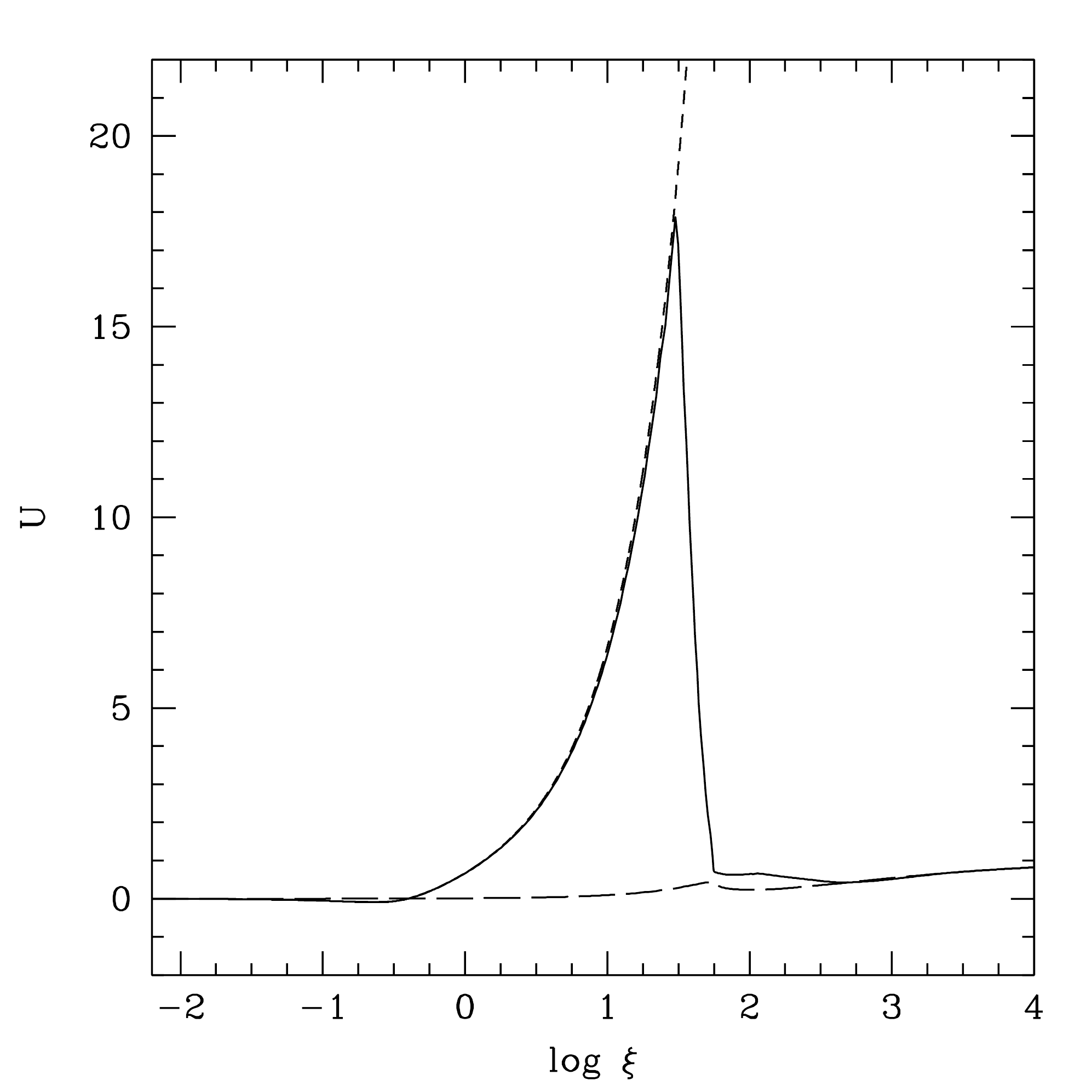}
\includegraphics[width=7cm]{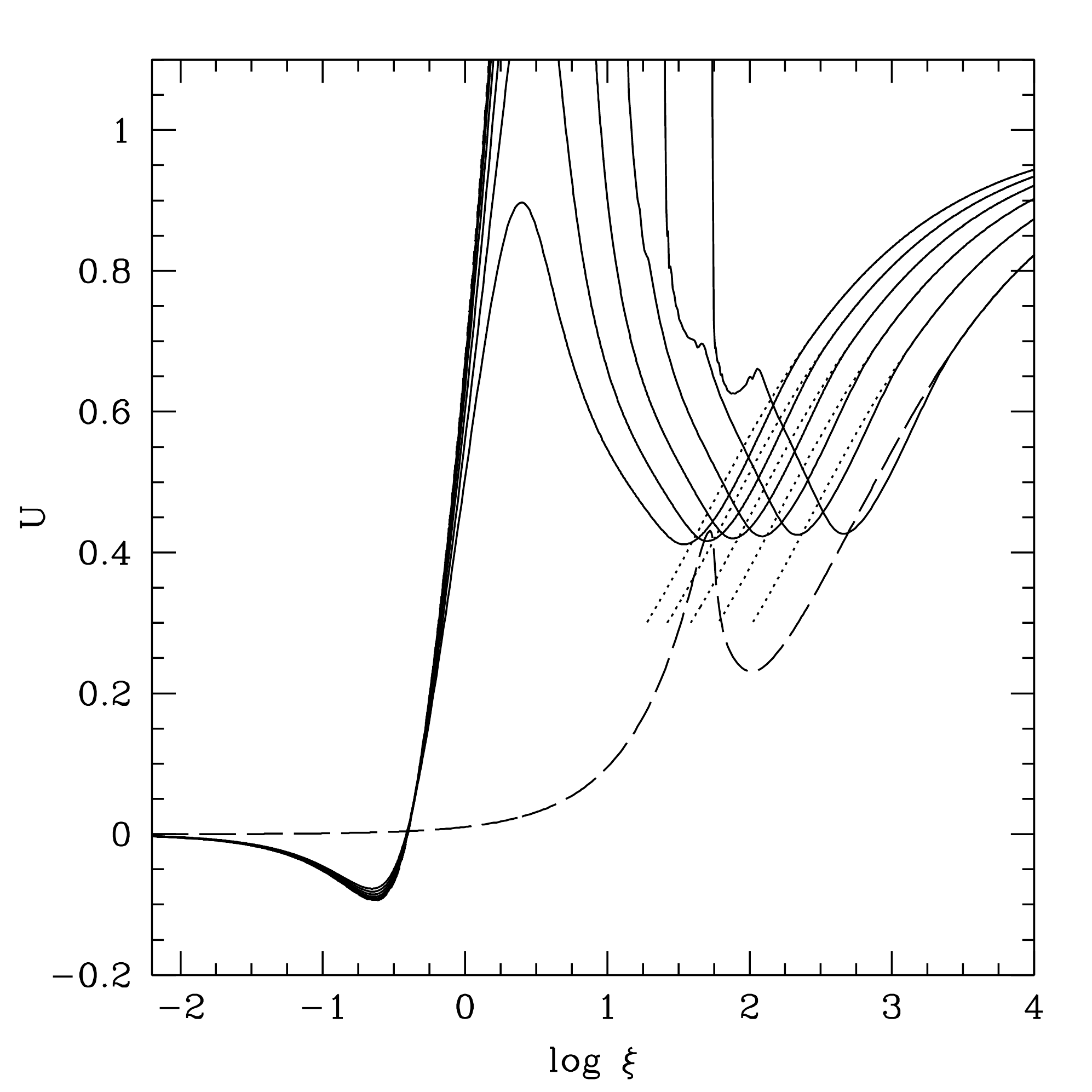}
 \caption{\label{fig.4} \small Simulation results for the velocity $U$ (from 
the same run as in figure \ref{fig.2}) plotted against the similarity 
coordinate $\xi$. The top left-hand frame shows curves for a succession of 
times during the intermediate state (the higher peaks corresponding to the 
later ones). The similarity solution is marked with the short-dashed curve. The 
top right-hand frame shows only the last of these time levels, but together 
with a long-dashed curve indicating the unperturbed FRW solution mapped onto 
this same space-time slice. The bottom frame is a zoom of the first one, 
looking in detail at the range $U \leq 1$. The long-dashed curve marks the same 
mapping of the FRW solution as before and the truncated dotted lines show part 
of the corresponding mappings for the earlier time levels. See text for further 
details.}
 \end{figure}

\begin{figure}[t!]
\centering
\includegraphics[width=7cm]{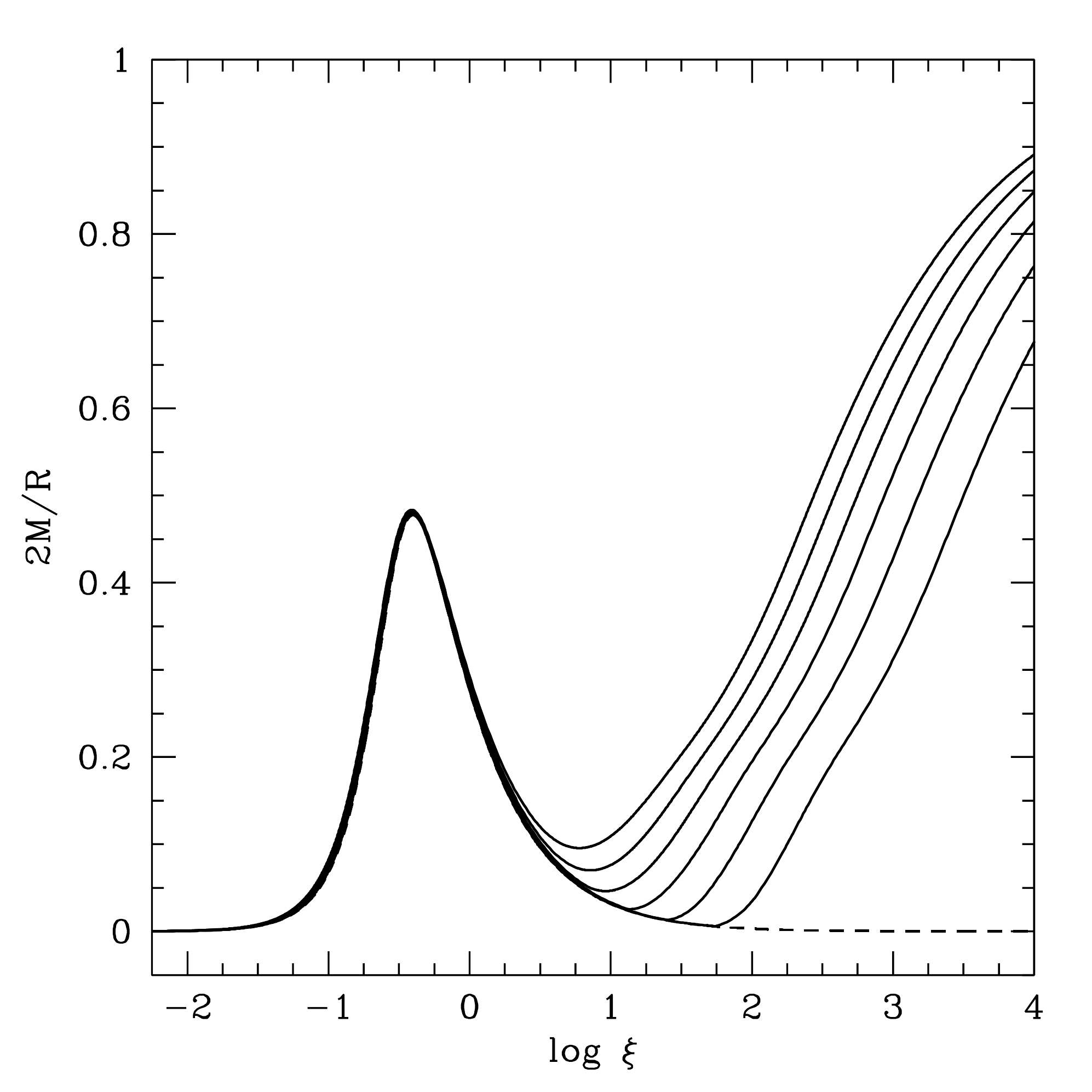}
\includegraphics[width=7cm]{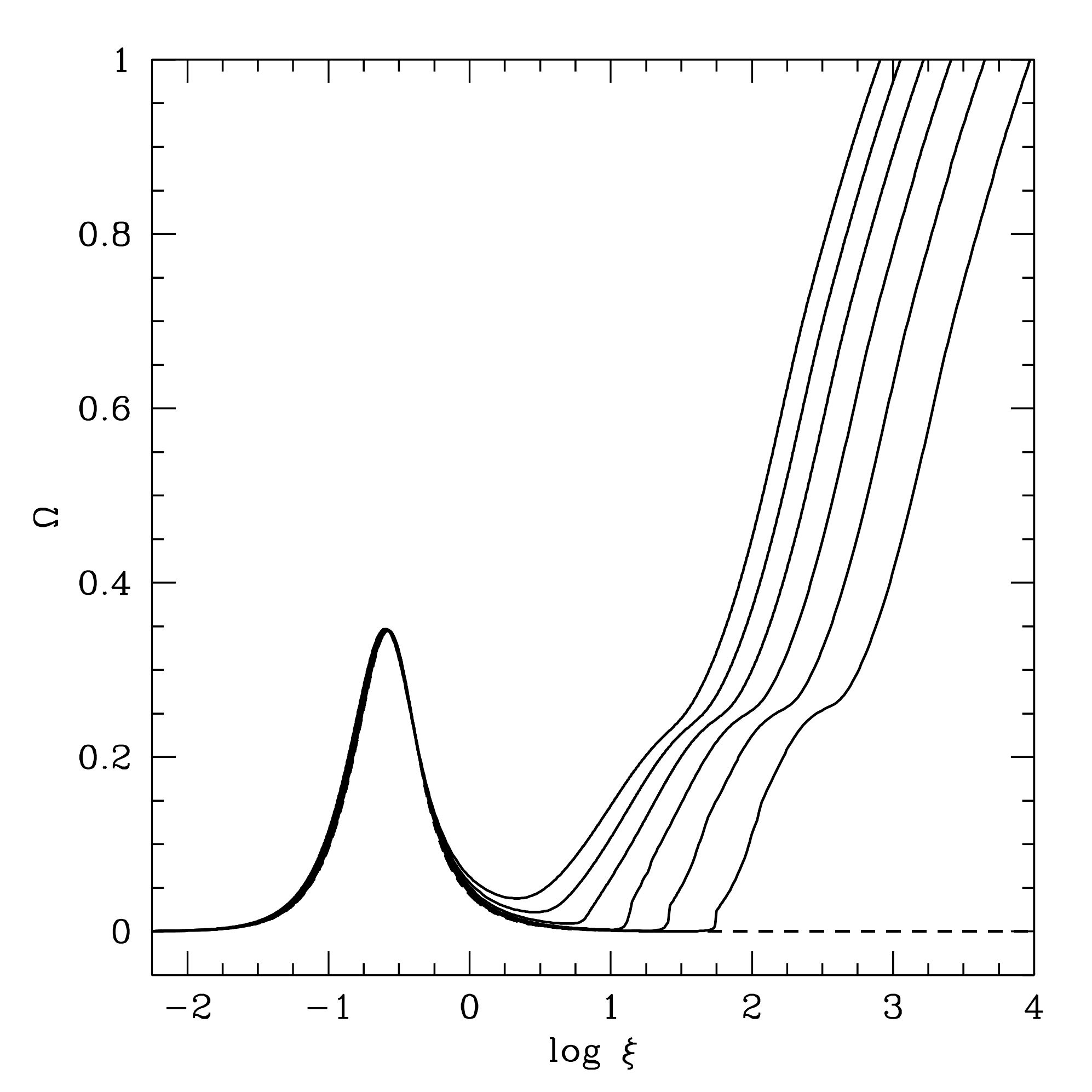}
 \caption{\label{fig.5} \small Corresponding simulation results (to those of 
the previous figure) for $2M/R$ ($=2\Phi$) and $\Omega$ ($=4\pi R^2 e$). The 
similarity solutions are again marked with the short-dashed curves.}
 \end{figure}

Figures \ref{fig.4} and \ref{fig.5}, which use data from the same run as in 
figure \ref{fig.2}, show in detail how the simulation results for the 
intermediate state approach the similarity solution. Quantities from the 
simulation are plotted against the similarity coordinate $\xi$ for a succession 
of times during the intermediate state (the first being for $t/t_H=5.072$ and 
the last being for $t/t_H=6.342$ which is just before $(2M/R)_{peak}$ starts to 
increase). Figure \ref{fig.4} shows three different views of the velocity $U$, 
with the bottom frame being a zoom of the top left-hand one, looking at the 
range of $U \leq 1$; figure \ref{fig.5} shows $2M/R$ ($=2\Phi$) and $\Omega$ 
($=4\pi R^2 e$). The later times correspond to the lower curves on the 
right-hand side of the plots and to the upper curves in the mid-region of the 
frames in figure \ref{fig.4}. The similarity solutions of figure \ref{fig.1} 
are shown here with short-dashed curves, but these are often covered. One can 
see the progressive approach of the simulation results towards the similarity 
solutions, with the range of the zone of agreement increasing with time. At the 
last time shown, the similarity solution is closely approximating the 
simulation results over all of the contracting region, where $U$ is negative, 
and also over the part of the surrounding region out to the maximum of $U$ (the 
peak of the relativistic wind which coincides with the minimum of the energy 
density in the evacuated region - see figure 4 of Paper 3). Beyond this, the 
simulation results diverge completely away from the similarity solution, 
eventually merging into the surrounding FRW universe. The top right-hand frame 
of figure \ref{fig.4} shows the last of the time levels seen in the previous 
frame, together with the short-dashed curve for the similarity solution and a 
long-dashed curve indicating the unperturbed FRW solution mapped onto this same 
space-time slice. (This curve has been obtained by calculating the local value 
of the Hubble constant $H = 1/2t_{\rm pr}$, where $t_{\rm pr}$ is proper time 
as measured by the local co-moving observers, and then taking $U = HR$.) In the 
bottom frame, with the expanded view for $U \leq 1$, the long-dashed curve 
again shows the mapping of the FRW solution onto the same space-time slice as 
before and the truncated dotted lines show part of the corresponding curves for 
the earlier time levels. Note that the strange form of the long-dashed curve 
arises because of the perturbation of the null slice onto which the FRW 
solution is being mapped. The join between the computed solution and the FRW 
solution in the outer region comes at the same locations, of course, also in 
the plots of figure \ref{fig.5} and excellent agreement is found there between 
the numerical and analytic results in all cases.

Regarding figure \ref{fig.4}, we should stress that the use of a logarithmic 
coordinate here has the effect of making features appear much more abrupt than 
they would with a standard linear coordinate. The almost vertical parts of the 
curves in the top frames are nowhere near to being shocks and correspond to 
smoothly-varying features when viewed on a linear scale (c.f. figure 4 of Paper 
3). The zoomed bottom frame has been primarily included so as to show clearly 
the transition to the FRW solution. However, it also reveals an interesting 
wave-like feature at the later times (seen in the solid curves), which we have 
then investigated in more detail. This turns out to result from interaction of 
the wind with the outer edge of the growing void and appears as a rather gentle 
space-variation when viewed on a linear scale.

\begin{figure}[t!]
\centering
\includegraphics[width=7cm]{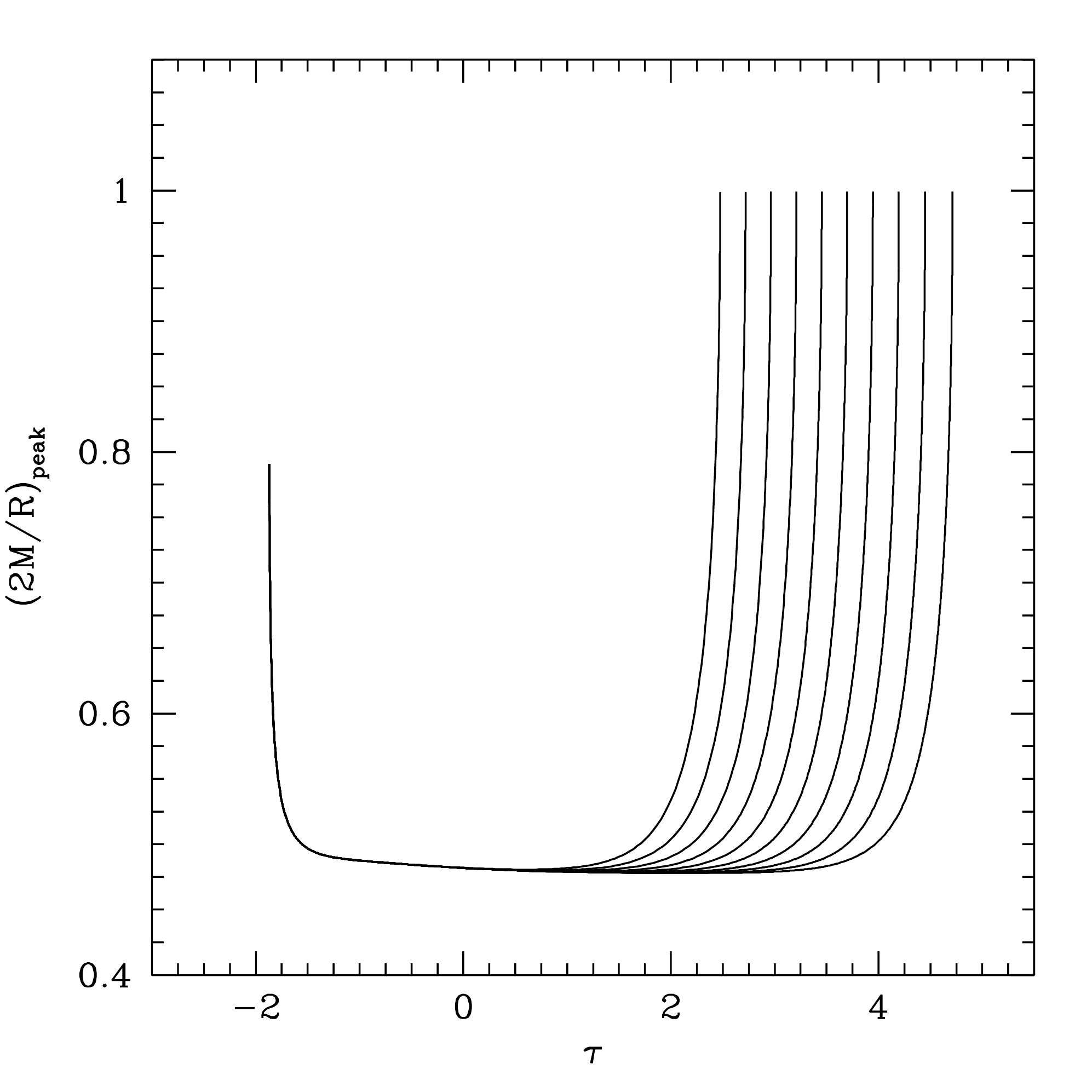}
\includegraphics[width=7cm]{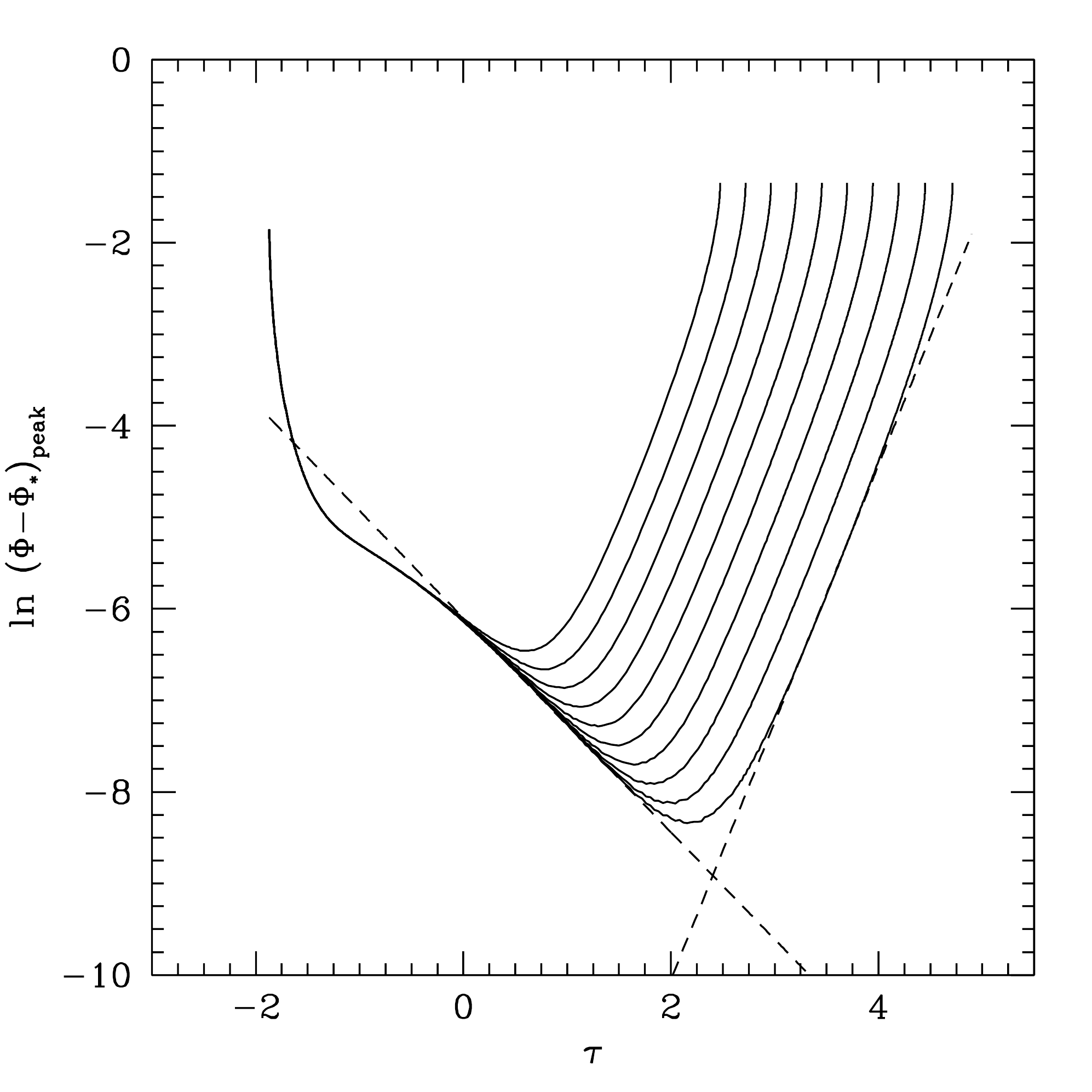}
\caption{\label{fig.6} \small Plots for investigating growing and decaying 
perturbation modes around the similarity solution. See text for details.}
 \end{figure}

From the right-hand frame of figure \ref{fig.2}, showing the intermediate state 
during which the numerical results are very close to those coming from the 
similarity solution (as seen in figures \ref{fig.4} and \ref{fig.5}), it is 
very clear that there is a systematic behaviour in the approach of the 
numerical solution to the self-similar one. There is also a systematic 
behaviour in the eventual departure from the similarity solution, leading 
towards black hole formation. In the earlier work (see the review 
\cite{Gundlach}), studies have been made of linear perturbations around the 
critical solution in terms of growing and decaying modes. In particular, the 
linear growing-mode index has been shown by Maison \cite{Maison} to be equal to 
$1/\gamma$. Following \cite{Maison} (but with some variations), it is 
convenient to write the growing-mode amplitude for our quantity $\Phi$ as
 \beq
\Phi(\xi,\tau) - \Phi_\star(\xi) \, \propto \, 
(\delta - \delta_c)e^{\lambda_0 \tau} \psi_0(\xi) \, ,
\eeq
 where $\Phi_\star(\xi)$ is the similarity solution and 
$\tau = -\ln[(t_c - t)/t_H]$. Note that increasing $\tau$ corresponds to    
increasing $t$, and that
 \beq
e^{\lambda_0 \tau} = [t_H/(t_c - t)]^{\lambda_0}
\eeq
 with $\lambda_0$ being the index. Similarly, one can write the decaying-mode 
amplitude as
 \beq
\Phi(\xi,\tau) - \Phi_\star(\xi) \, \propto \, 
e^{\lambda_1 \tau} \psi_1(\xi) \, ,
\eeq
 (with no dependence on $(\delta - \delta_c)$ being expected). We apply these 
to the peak value of $\Phi$, i.e. $(2M/R)_{\rm peak}/2$. In figure \ref{fig.6}, 
we first plot (in the left frame) simulation results for $(2M/R)_{peak}$ as 
functions of $\tau$, coming from a succession of runs with different values of 
$(\delta - \delta_c)$ equally spaced in the log (decreasing successively by 
factors of $2$ moving from left to right). The $\log_{\,e}$ of the growing mode 
amplitude at the peak $(\Phi - \Phi_\star)_{\rm peak}$ is then plotted against 
$\tau$ in the right-hand frame. From the equal spacing of the curves in the 
rising part, one can immediately see the expected linear dependence on $(\delta 
- \delta_c)$ for the growing mode. By making straight-line fits to the most 
linear segments of the growing and decaying parts of the curves, one can read 
off values of $\lambda_0$ for the growing mode and $\lambda_1$ for the decaying 
mode. (We have marked our straight-line fits with the dashed lines.) We find 
$\lambda_0 = 2.81$ for the growing mode and $\lambda_1 = -1.17$ for the 
decaying mode. Again following Maison \cite{Maison}, we anticipate that 
$1/\lambda_0$ should be equal to $\gamma$, the index of the scaling law for 
$M_{BH}$. In fact, we have that $1/\lambda_0 = 0.356$ which compares with 
$\gamma = 0.3558$ as given by Maison for $w = 1/3$ and our value of $\gamma = 
0.357$ as given in Paper 3. We should stress, however, that while $\lambda_0 = 
2.81$ was genuinely our best fit, we do not believe that it should be trusted 
to better than about $\pm 2$ in the last digit.

\section{Investigation of the effects of varying $w$ and the shape of the 
initial perturbation}

In the literature on critical collapse, there has been discussion of the 
consequences of using ``perfect fluids'' having equations of state of the form 
$p = we$, with $w$ taking values covering the range from $0$ to $1$ 
\cite{Maison,Choptuik2}. Our own interest is strictly related to the 
possibility that critical collapse might actually occur under realistic 
circumstances, and a particular context for this arises in the case of the 
radiation fluids which dominated the universe at early times and could be 
reasonably approximated with an equation of state of the form $p = we$ with $w 
= 1/3$. Investigation of similar equations of state with $w \ne 1/3$ may be of 
interest in relation to this, for giving some indication of the effects of 
softening of the equation of state at the time of phase transitions, or 
stiffening due to some non-standard interactions. One needs to proceed with 
caution, however.

Standard simple fluids have pressure proportional to internal energy density, 
not to the {\it total} energy density (including also the rest-mass energy of 
the constituent particles). The form $p = we$ is relevant in the limit where 
the rest-mass energy becomes negligible compared with the kinetic energy of 
random motions (as for a radiation fluid). Using $p = we$ for a perfect fluid 
with positive values of $w$ different from $1/3$ has doubtful physical 
motivation.

\begin{figure}[t!]
\centering
\includegraphics[width=7cm]{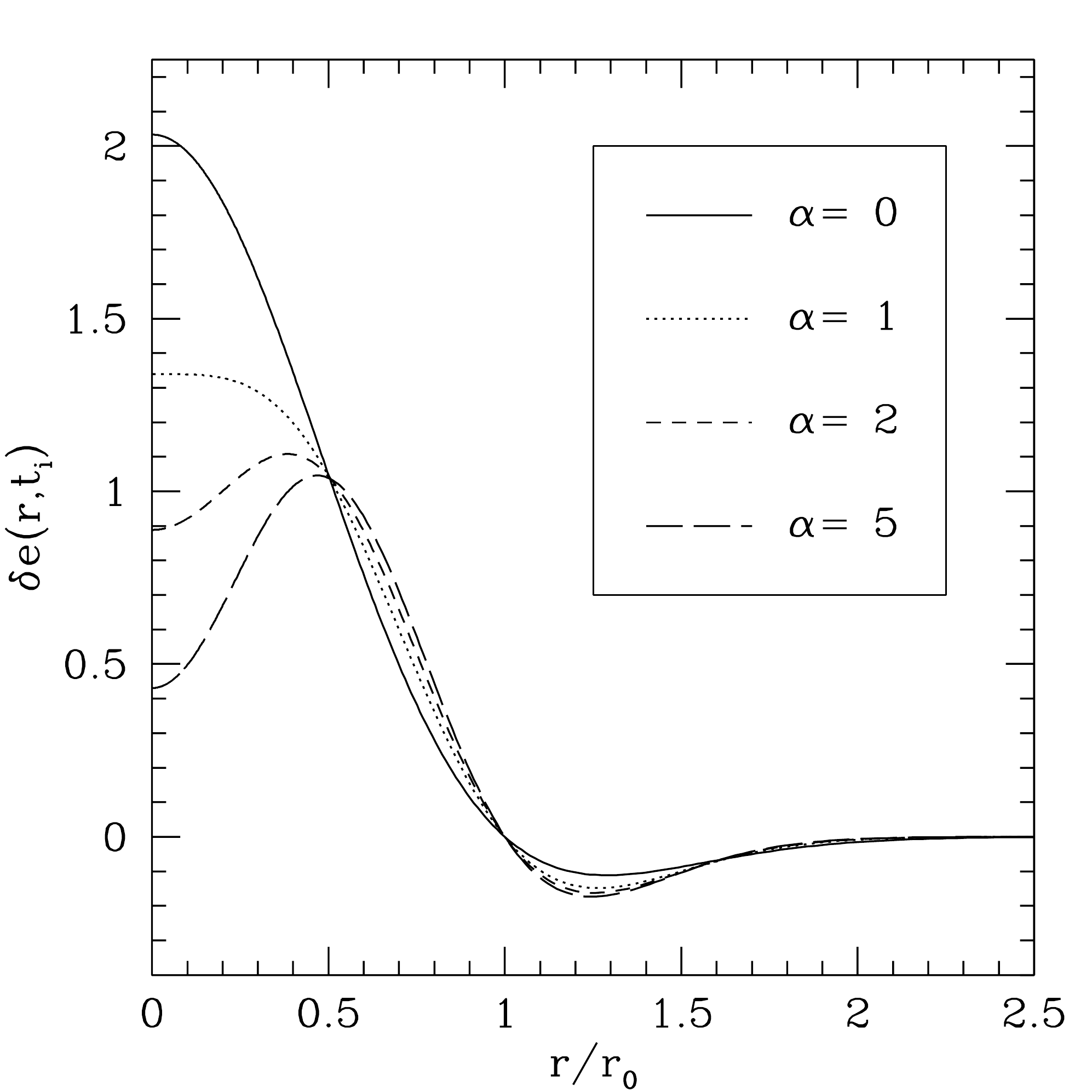}
 \caption{\label{fig.7} \small Different shapes of the energy-density 
perturbation obtained with different values of $\alpha$, for the particular 
case $w = 1/3$ and $\delta=\delta_c$.}
 \end{figure}

Having made that caveat, however, we now look at cases with $w \ne 1/3$ with 
the aim mentioned above. We use exactly the same approach as that used for $w = 
1/3$, even though there is some artificiality in doing this because of having 
the whole universe following an unusual equation of state and imposing initial 
conditions under those circumstances. If one were really considering in detail 
possible PBH formation with a variant equation of state, one would need a more 
sophisticated set-up than our present one. However, our aim here is just to get 
some general indications about the effect of varying $w$. We have studied a 
range of values for $w$ between $0.01$ and $0.6$, checking on how the values of 
$\gamma$ and $\delta_c$ for the scaling laws varied with $w$ and also with 
varying the shape of the initial perturbation. Regarding the hydrodynamics of 
the process: as $w$ is increased, the features of the relativistic wind and the 
opening up of the void become progressively more extreme, and this eventually 
becomes very challenging for the adaptive scheme. It is because of this that we 
did not use values of $w > 0.6$. For the variant perturbation shapes, we used 
the forms of curvature profile given by equation (\ref{curvature_profile}) with 
$\alpha \ne 0$. The corresponding perturbations of the energy density, given by 
equation (\ref{de2}), are shown in figure \ref{fig.7} for the particular case 
$w=1/3$ and $\delta=\delta_c$, as an indication of the general behaviour. 
Varying $\alpha$ in the range between $0$ and $1$ gives a centrally-peaked 
perturbation shape, while for $\alpha$ greater than $1$ it is off-centred.

\begin{figure}[t!]
\centering
\includegraphics[width=7cm]{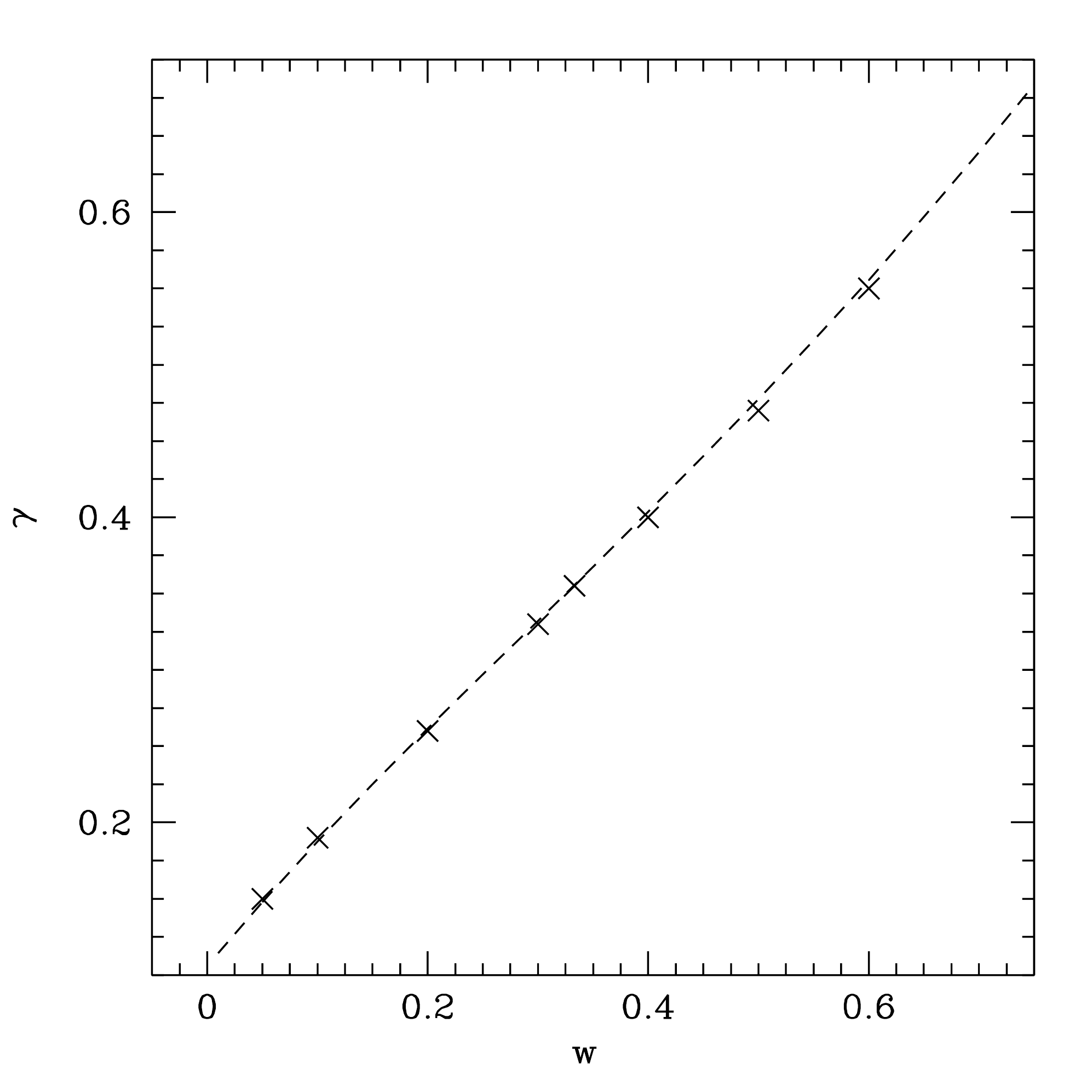}
\includegraphics[width=7cm]{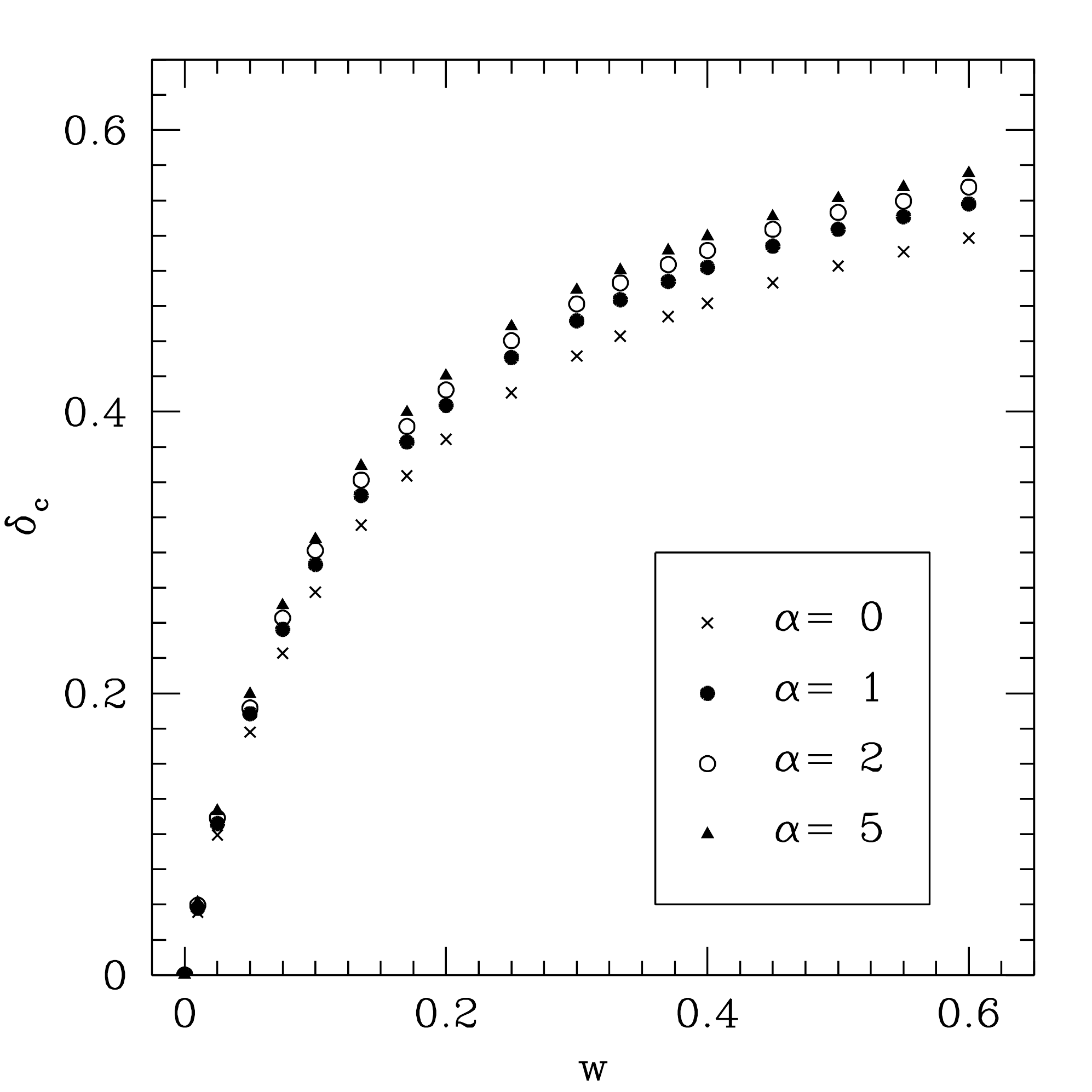}
 \caption{\label{fig.8} \small The left-hand plot shows the behaviour of 
$\gamma$ as a function of $w$ (with the dashed line indicating the 
corresponding results obtained semi-analytically by Maison \cite{Maison}). The 
right-hand plot shows the behaviour of $\delta_c$ as a function of $w$, as well 
as the variations depending on the shape parameter $\alpha$ (these variations 
are negligible in the left-hand plot and so are not shown there).}
 \end{figure}

With all of the cases which we studied ($0.01 \le w \le 0.6$ and $0 \le \alpha 
\le 5$) we found good scaling laws extending down to the smallest values of 
$(\delta-\delta_c)$ for which we were able to make satisfactory calculations. 
The intermediate state and approximate similarity solution were clearly seen in 
all cases, with the value of $(2M/R)_{peak}$ decreasing with decreasing $w$. 
Figure \ref{fig.8} shows our best-fit values of $\gamma$ and $\delta_c$ as 
functions of $w$. In the case of $\gamma$, the results obtained for different 
values of $\alpha$ are indistinguishable on the scale of the plot, while for 
$\delta_c$ small variations are seen. The results for $\gamma$ are in good 
agreement with those previously obtained semi-analytically by Maison 
\cite{Maison} (which are indicated here with the dashed line). Within the range 
shown, there is a roughly linear behaviour which is consistent with the limit 
of $\gamma\rightarrow0.106$ for $w\rightarrow0$ obtained by Snajdr 
\cite{Snajdr}. Regarding $\delta_c$, although the change with $\alpha$ does not 
seem to be very large, it could be cosmologically relevant because the PBH mass 
spectrum is very sensitive to the precise value of $\delta_c$. It is therefore 
important to establish the connection between different inflationary models and 
the probabilities for different initial perturbation shapes, and more 
investigation of this should be made in future for getting a better 
understanding of the possible cosmological impact of PBHs. The plot indicates 
that among the profiles studied here, the simple Mexican hat (with $\alpha = 
0$) gives the lowest value of $\delta_c$ and so the highest probability of 
forming PBHs. The relation between $\delta_c$ and $w$ confirms that any epochs 
in the early universe when the equation of state softens should be favourable 
for enhancing PBH production. One example of this could be the QCD phase 
transition, which has previously been discussed in connection with a PBH model 
for MACHOs \cite{Jedamzik3}.

Finally we want to comment here on one of our earlier results, in Paper 1, 
where we investigated critical collapse for a radiative fluid also in the 
presence of a cosmological constant $\Lambda$. At the time of Paper 1, our code 
was not able to get very close to the critical limit because of not yet having 
the AMR, and the lowest value of $(\delta-\delta_c)$ that we were able to treat 
was around $10^{-3}$. In that regime we found that the presence of a 
cosmological constant was affecting the scaling law, giving a change in the 
value of $\gamma$ in the sense of decreasing it with increasing (positive) 
$\Lambda$. Now, with the present AMR code, we have made similar calculations 
going down to much smaller values of $(\delta-\delta_c)$ and have seen that 
this change in $\gamma$ disappears as one gets closer to $\delta_c$ (i.e. the 
gradients of the two scaling laws converge to the same value). The reason for 
this is clear: if $\delta$ is very close to $\delta_c$, the mass of the black 
hole is small and the fluid densities involved in forming it are high. Under 
these circumstances, the energy density related to the cosmological constant 
becomes negligible in comparison with that of the collapsing fluid. This makes 
the gradient of the scaling law converge to the same value as without the 
cosmological constant when $(\delta-\delta_c)$ is sufficiently small. However, 
the cosmological constant still makes a relevant (and growing) contribution on 
larger scales where the fluid density is lower.



\section{Conclusions}
 Following on after our previous work investigating primordial black hole 
formation during the radiative era of the early universe, we have here 
investigated further the critical nature of the collapse in this context, 
focusing on the intermediate state, which appears in the case of perturbations 
close to the critical limit. We have examined the extent to which this follows 
a similarity solution, deriving and solving the set of equations describing a 
self-similar solution within the same foliation used for our simulations of 
black hole formation. This is a key issue, because the presence of a similarity 
solution is seen as an important characteristic feature of the general 
phenomenon of critical collapse. Also, we have presented results from 
calculations where the equation of state parameter $w$ was allowed to take 
constant values different from the radiation value of $1/3$, with the aim of 
gaining further insight into our main case of interest, and have studied the 
effect of using different perturbation shapes. Our calculations have been made 
using a purpose-built Lagrangian AMR code, starting with initial supra-horizon 
scale perturbations of a type which could have come from inflation and then 
following self-consistently both the formation of the black hole and the 
continuing expansion of the universe.

From our simulation results, we have found that the similarity solution does 
indeed emerge in this context as an attractor solution approached during the 
intermediate state. However, for runs with perturbations whose amplitude 
$\delta$ is just above the critical value $\delta_c$, we observe it arising 
together with decaying and growing perturbation modes. First, the decaying mode 
is seen operating as the similarity solution is approached during the 
intermediate state but then the growing mode takes over, leading away from the 
intermediate state and the similarity solution, towards black hole formation. 
As expected, the index of the growing mode is found to be closely equal to 
$1/\gamma$, where $\gamma$ is the exponent of the scaling law for the 
black-hole mass. During the time of the intermediate state, the range over 
which the similarity solution gives a good approximation to the simulation 
results becomes progressively larger, eventually extending over all of the 
contracting region and part of the surrounding evacuated region, up to the 
maximum in the velocity of the relativistic wind, corresponding to the deepest 
point of the surrounding semi-void. Further out, the simulation results diverge 
completely away from the similarity solution, transitioning onto the 
surrounding expanding universe. From our simulations for cosmological-type 
perturbations using values for $w$ different from $1/3$, we have found that in 
every case, scaling-law behaviour persists down to the smallest values of 
$(\delta - \delta_c)$ for which we were able to make satisfactory calculations, 
with the values obtained for the exponent $\gamma$ being in almost perfect 
agreement with those obtained previously by other authors. The critical 
threshold amplitude $\delta_c$, the intermediate state compactness 
$(2M/R)_{peak}$ and the scaling-law exponent $\gamma$ all vary with $w$ in a 
way which is easily understandable in view of $w$ being the ratio between the 
pressure and energy density of the fluid.



\ack This work has been partly carried out within the award ``Numerical 
analysis and simulations of geometric wave equations'' made under the European 
Heads of Research Councils and European Science Foundation EURYI (European 
Young Investigator) Awards scheme, supported by funds from the Participating 
Organizations of EURYI and the EC Sixth Framework Programme. We gratefully 
acknowledge helpful discussions, during the course of this work, with a number 
of colleagues including Carsten Gundlach, Alexander Polnarev, Bernard Carr, Ian 
Hawke and Karsten Jedamzik, and IM thanks SISSA for giving him use of their 
facilities.


\end{document}